\documentclass[11pt]{article}
\usepackage[a4paper]{geometry}

\usepackage[inline]{enumitem} % allowed
\usepackage{mathtools} % allowed
\usepackage{xspace} % allowed
\usepackage{bm} % allowed

\usepackage{amssymb}
\usepackage{amsthm}

\usepackage{hyperref} % allowed
\usepackage[capitalise]{cleveref} % not on the list !?
\hypersetup{
    colorlinks=true,
    citecolor=blue,
    linkcolor=blue,
    filecolor=blue,
    urlcolor=blue,
}
\usepackage{longtable}

\usepackage{textpos}

\newtheorem{theorem}{Theorem}

\newtheorem{corollary}[theorem]{Corollary}
\newtheorem{lemma}[theorem]{Lemma}

\newtheorem{proposition}[theorem]{Proposition}
\newtheorem{definition}[theorem]{Definition}

\newtheorem{claim}{Claim}
\newcommand{\wh}{\widehat}
\newenvironment{proof*}[1][\proofname]{
  
  \begin{proof}[#1]}{\end{proof}}

\usepackage{thmtools}
\usepackage{thm-restate}

\def\Gajarsky{Gajarsk\'{y}\xspace}

\newcommand{\first}{\text{first}}

\newcommand{\from}{\colon}
\newcommand{\tto}{\rightsquigarrow}

\newcommand{\set}[1]{\{#1\}}
\newcommand{\setof}[2]{\set{#1\mid#2}}
\def\phi{\varphi}
\def\cal{\mathcal}
\def\N{\mathbb N} %mathbb?

\def\epsilon{\varepsilon}
\def\eps{\varepsilon}
\renewcommand{\subset}{\subseteq}
\renewcommand{\setminus}{-}
\renewcommand{\le}{\leqslant}
\renewcommand{\ge}{\geqslant}
\renewcommand{\leq}{\leqslant}

\newcommand{\CC}{\cal C}
\newcommand{\DD}{\cal D}

\newcommand{\BB}{\cal B}
\newcommand{\trans}[1]{\mathsf{#1}}

\newcommand\wreach{{\rm WReach}}

\newcommand\wcol{{\rm wcol}}
\newcommand{\leaves}{\mathrm{Leaves}}

\newcommand{\FO}{\text{\rm{FO}}\xspace}

\newcommand{\Oof}{\mathcal O}
\newcommand{\Oh}{\Oof}

\newcommand{\tup}{\bar}
\newcommand{\tp}{\textnormal{tp}}

\newcommand{\struct}[1]{\mathbf{#1}}

\newcommand{\Ff}{\cal F}

\newcommand{\EE}{\mathcal{E}}

\newcommand{\BS}{B_r^{\mathrm{sep}}}

\newcommand{\rep}{\mathrm{rep}}

\title{Treelike Decompositions\\ for Transductions of Sparse Graphs\thanks{This work is a part of projects LIPA (JG, SK) and BOBR (JG, MP, SzT) that have received funding from the European Research Council (ERC) under the European Union’s Horizon 2020 research and innovation programme (grant agreements No. 683080 and 948057, respectively).}}
\author{Jan Dreier\thanks{TU Wien, Austria (\texttt{dreier@ac.tuwien.ac.at})} \and  Jakub \Gajarsky\thanks{Institute of Informatics, University of Warsaw, Poland (\texttt{jakub.gajarsky@mimuw.edu.pl})} \and Sandra Kiefer \thanks{RWTH Aachen University, Germany (\texttt{kiefer@informatik.rwth-aachen.de})}\and Michał Pilipczuk\thanks{Institute of Informatics, University of Warsaw, Poland (\texttt{michal.pilipczuk@mimuw.edu.pl})} \and Szymon Toruńczyk\thanks{Institute of Informatics, University of Warsaw, Poland (\texttt{szymtor@mimuw.edu.pl})}}
\begin{document}

\maketitle
\thispagestyle{empty}

\begin{textblock}{20}(-1.9, 6)
	\includegraphics[width=40px]{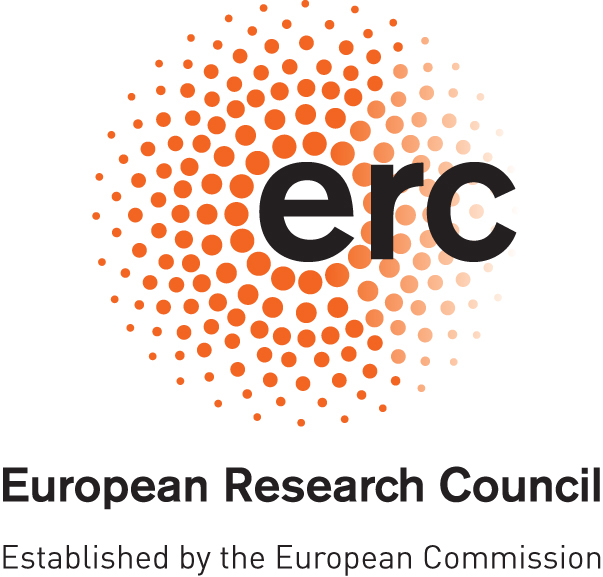}%
\end{textblock}
\begin{textblock}{20}(-2.27, 6.4)
	\includegraphics[width=70px]{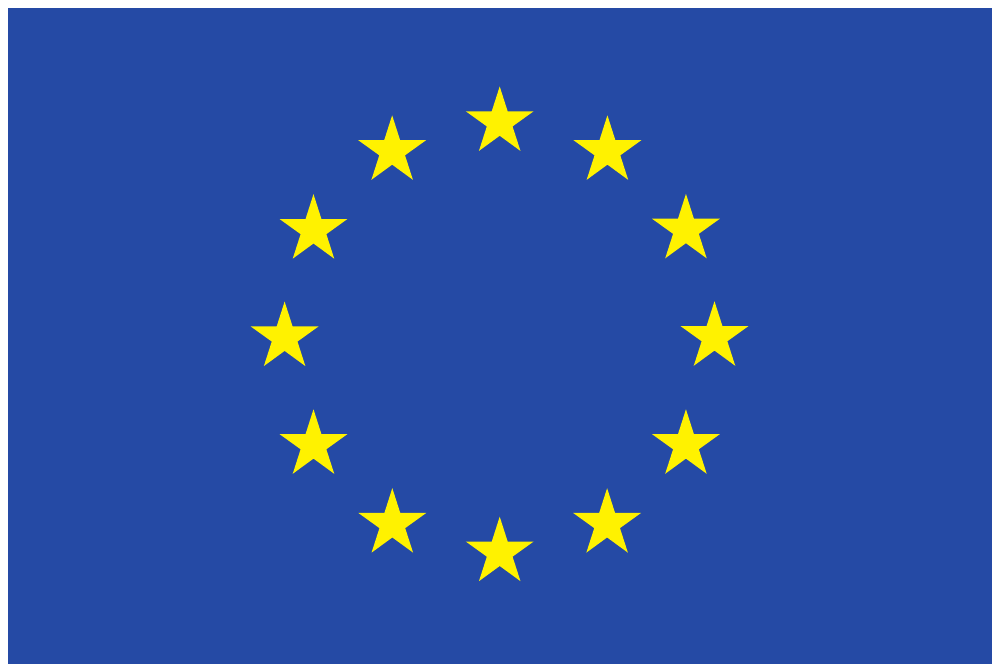}%
\end{textblock}

\begin{abstract}
 We give new decomposition theorems for classes of graphs that can be transduced in first-order logic from classes of sparse graphs\,---\,more precisely, from classes of bounded expansion and from nowhere dense classes. In both cases, the decomposition takes the form of a single colored rooted tree of bounded depth where, in addition, there can be links between nodes that are not related in the tree. The constraint is that the structure formed by the tree and the links has to be sparse. Using the decomposition theorem for transductions of nowhere dense classes, we show that they admit low-shrubdepth covers of size $\Oh(n^\eps)$, where $n$ is the vertex count and $\eps>0$ is any fixed~real. This solves an open problem posed by \Gajarsky et al.\ (ACM TOCL '20) and also by Briański et al.\ (SIDMA '21).\end{abstract}

\newpage

%!TEX root = main.tex
\section{Introduction}\label{sec:intro}

We study classes of {\em{structurally sparse graphs}}, i.e., graphs interpretable in sparse graphs using first-order logic ($\FO$). The ultimate goal of this line of research, pursued earlier in~\cite{8576442,GajarskyKNMPST20,Dreier21}, is to obtain algorithmic results for structurally sparse graphs by lifting methods used in the sparse setting. While this seems currently out of reach, the present focus --- and the objective of this work --- is to obtain a purely combinatorial description of structurally sparse graphs, which might get us closer to the desired algorithmic insights.

We mainly focus on two notions of sparsity\,---\,nowhere denseness and bounded expansion, defined as follows.
A graph $H$ is a {\em{depth-$d$ minor}} of a graph $G$ if one can obtain $H$ from a subgraph of $G$ by contracting mutually disjoint connected subgraphs of radius at most $d$.
A class of graphs $\CC$ is {\em{nowhere dense}} if, for every $d\in \N$, there is a uniform bound $t(d)\in \N$ on the maximum size of a complete graph that is a depth-$d$ minor of a graph from $\CC$. More restrictively, $\CC$ has {\em{bounded expansion}} if, for every $d\in \N$, there is a uniform bound $c(d)\in \N$ on the maximum average degree of a depth\nobreakdash-$d$ minor of a graph from $\CC$. Clearly, every class of bounded expansion is nowhere dense, but there exist classes that are nowhere dense and have unbounded expansion~\cite{sparsity}.

The two definitions form the foundations of {\em{Sparsity}}, the theory of classes of sparse graphs, which was initiated by Ne\v{s}et\v{r}il and Ossona de Mendez~\cite{sparsity}. This theory has developed rapidly over the last 15 years and has provided a wealth of combinatorial tools for the treatment of sparse graphs, or, more precisely, graphs belonging to any fixed class that is either of bounded expansion or nowhere dense. Since the two notions generalize multiple contemporary concepts of sparsity\,---\,like having bounded maximum degree, bounded treewidth, or excluding a fixed (topological) minor\,---\,the obtained techniques have versatile applications and far-reaching consequences~(see e.g. \cite{sparsity,notes}). 

One of the great successes of the Sparsity program is a uniform explanation of tractability of model-checking $\FO$ in monotone (that is, subgraph-closed) graph classes. More precisely, Grohe et al.~\cite{GroheKS2017} showed that for every fixed nowhere dense class $\CC$, every $\FO$ sentence $\varphi$, and every $\eps>0$, there is an algorithm that decides whether $\varphi$ holds in a given graph $G\in \CC$ in time $\Oh(n^{1+\eps})$, where $n$ is the number of vertices of $G$. On the other hand, once $\CC$ is not nowhere dense and is monotone, no model-checking algorithm with running time $\Oh(n^c)$
for any constant $c$ should be expected under standard assumptions from Parameterized Complexity. Conceptually, this means that as far as monotone classes are concerned, nowhere denseness exactly delimits the region of algorithmic tractability of model-checking $\FO$ on graphs. We remark that an earlier result due to Dvo\v{r}\'ak et al.~\cite{DvorakKT13} established an $\Oh(n)$-time algorithm for model-checking $\FO$ on classes of bounded expansion, and both the results~\cite{GroheKS2017} and~\cite{DvorakKT13} naturally extend to the setting of relational structures when assuming the sparsity of their Gaifman graphs.

Together with the host of earlier and subsequent advances, the results from~\cite{GroheKS2017} and~\cite{DvorakKT13} provide a robust toolbox for working with $\FO$ on sparse graphs. It is interesting to investigate to what extent this toolbox can be used beyond the context of sparsity, or more precisely, on graphs that might be dense but are otherwise structurally well-behaved. A first step would be to consider classes of graphs that can be interpreted using $\FO$ in sparse graphs, but already here,  severe complications arise.

We use {\em{transductions}} to formalize the concept of $\FO$-defi\-nable graph transformations. A (simple) $\FO$ transduction is a non-deterministic mechanism that inputs a graph, outputs another graph, and consists of the following three steps: (1) non-deterministically choose a coloring of the vertex set using a bounded number of colors; (2) interpret a new adjacency relation using a fixed $\FO$ formula $\psi(x,y)$; and (3) output any induced subgraph of the obtained graph\footnote{It is also allowed to copy the universe a bounded number of times, but this is immaterial for this overview; see Section~\ref{sec:prelims} for a formal definition.}. We say that a class $\DD$ can be {\em{transduced}} from a class $\CC$ if there is a fixed transduction $\trans T$ such that every $G\in \DD$ can be obtained by applying $\trans T$ on some $H\in \CC$. Following terminology from~\cite{GajarskyKNMPST20}, classes that can be transduced from classes of bounded expansion are said to have {\em{structurally bounded expansion}}, while classes that can be transduced from nowhere dense classes are {\em{structurally nowhere dense}}.

So far we have a rough combinatorial description of structurally bounded expansion classes. First, Gajarsk\'y et al.~\cite{GajarskyKNMPST20} proved that they coincide with classes that admit {\em{low shrubdepth covers}} of bounded size. Informally speaking, a graph has bounded shrubdepth if it admits a decomposition, called a {\em{connection model}}, which is a clique expression (like in the definition of cliquewidth) of bounded depth\footnote{Model-theoretically, classes of bounded shrubdepth are exactly those that can be transduced from classes of bounded-depth trees. Shrubdepth was introduced in~\cite{GanianHNOM19} and can be seen as a dense analogue of treedepth.}. 

\begin{theorem}[\cite{GajarskyKNMPST20}]\label{thm:lsd-be}
 A class $\DD$ of graphs  has structurally bounded expansion if and only if the following condition holds.
  For every $p\in \N$, there is a constant $m=m(p)$ such that for every graph $G\in \DD$, one can find a family $\Ff(G)$ of vertex subsets of $G$ with $|\Ff(G)|\leq m$ and the following properties:
 \begin{itemize}
  \item for every $X\subseteq V(G)$ with $|X|\leq p$, there is $A\in \Ff(G)$ such that $X\subseteq A$; and
  \item the class $\{G[A]\colon G\in \DD, A\in \Ff(G)\}$ of induced subgraphs has bounded shrubdepth.
 \end{itemize}
\end{theorem}
The condition in the theorem is commonly abbreviated 
to saying that $\DD$ admits \emph{low shrubdepth covers}.

\medskip

More recently, Dreier~\cite{Dreier21} proposed two different structural characterizations of classes of structurally bounded expansion, via {\em{lacon decompositions}} and {\em{shrub decompositions}}. Without going into details, a shrub decomposition represents a given graph through a sparse graph model, colored with a bounded number of colors. The adjacency between a pair of vertices $u,v$ in the original graph can be deduced in the model by looking at the pair of colors of $u$ and $v$ and the distance between $u$ and $v$. Lacon decompositions have a more complicated definition and are based on local separability properties in sparse graphs, expressed through {\em{weak coloring numbers}}. All in all, compared to Theorem~\ref{thm:lsd-be}, both lacon and shrub decompositions provide a single, global decomposition of graphs belonging to a fixed class of bounded expansion.

As for structurally nowhere dense classes, not much is known. In particular, given Theorem~\ref{thm:lsd-be} and other results in the theory~(cf.~\cite[Theorem 13.1]{sparsity}), it is natural to conjecture that if a class $\DD$ is structurally nowhere dense, then it admits low shrubdepth covers in the sense of Theorem~\ref{thm:lsd-be}, but with the cardinality of the cover bounded by $\Oh_{\DD,p,\eps}(n^\eps)$, where $n$ is the vertex count of $G$ and $\eps>0$ is any fixed real\footnote{The $\Oh_{\tup x}(\cdot)$ notation hides factors depending solely on $\tup x$.}. Unfortunately, the proof of Theorem~\ref{thm:lsd-be} does not lift to the structurally nowhere dense case, because it is based on a quantifier-elimination procedure for bounded-expansion classes, and no such procedure is known for nowhere dense classes. Consequently, the question has remained open; it was implicitly asked in~\cite{GajarskyKNMPST20} and repeated in~\cite{BrianskiMPS21}.
There has been some work obtaining weaker decomposition theorems and structural properties for structurally nowhere dense classes, see~\cite{BrianskiMPS21,NesetrilMPZ20}. In particular, the aforementioned question about low shrubdepth covers was confirmed for $p=1$ and the power graph construction (a specific transduction) in~\cite{NesetrilMPZ20}. However, all in all, robust decomposition notions for structurally nowhere dense  classes haven't been found.

\paragraph*{Our contribution.} We introduce new decomposition concepts\,---\,{\em{bushes}} and {\em{quasi-bushes}}\,---\,which apply to classes with structurally bounded expansion and structurally nowhere dense classes, respectively. More precisely, if $\DD$ has structurally bounded expansion, then every graph $G\in \DD$ can be described by a bush of bounded depth, and all these bushes form a class of bounded expansion. An analogous statement applies to structurally nowhere dense classes and quasi-bushes, with the caveat that the obtained class of quasi-bushes is not necessarily nowhere dense, strictly speaking, but enjoys quantitative key properties of nowhere dense classes (e.g. weak coloring numbers bounded by $\Oh_{\DD,d,\eps}(n^\eps)$). Using quasi-bushes, we answer in affirmative the aforementioned question about the existence of small low-shrubdepth covers in structurally nowhere dense classes.

The definition of a bush stems from the concept of connection models, which is the decomposition notion underlying shrubdepth. A {\em{connection model}} for a graph $G$ is a tree $T$ labeled with a bounded number of labels such that:
\begin{itemize}
 \item the vertices of $G$ are the leaves of $T$; and
 \item for two vertices $u,v\in V(G)$, whether $u$ and $v$ are adjacent in $G$ depends only on the labels of $u$ and $v$ in~$T$, and the label of the lowest common ancestor of $u$ and $v$ in $T$.
\end{itemize}

\begin{figure}[t!]\centering
% \begin{minipage}[t]{.49\textwidth}
  \includegraphics[page=1,width=0.7\textwidth]{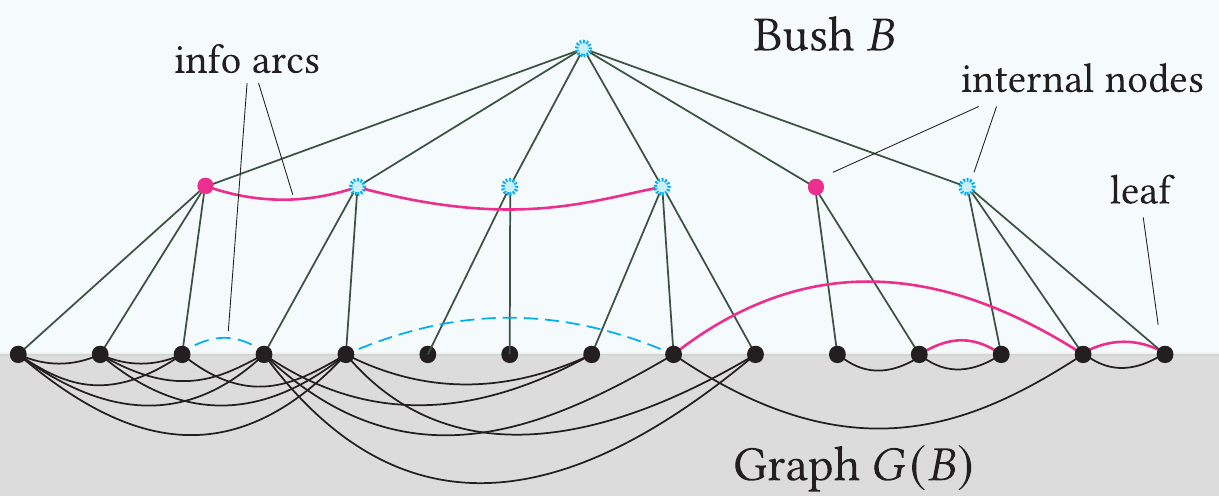}
  \caption{A bush $B$ and the graph $G(B)$ defined by it.
  It is a tree equipped with \emph{info arcs} that may connect nodes at the same depth. 
  In this bush, the leaves use one label only,
  which turns out to be always sufficient. 
  There are two labels for the info arcs: red and blue (dashed). Each internal node is equipped with an \emph{info loop} whose label is indicated by the color of the node. The vertices of $G(B)$ are the leaves of $B$, and two vertices $u,v$ are adjacent in $G(B)$ if and only if the lowest info arc (or info loop) above $u$ and $v$ is~red.
  }
  \label{fig:bush}
% \end{minipage}
  % \hfill
  % \begin{minipage}[t]{.49\textwidth}  
  % \end{minipage}
\end{figure}
A class $\DD$ has bounded shrubdepth if there are some $d,\ell\in \N$ such that every $G\in\DD$ has a connection model of depth $d$ using $\ell$ labels.

The purpose of our bushes is to characterize classes with structurally bounded expansion in a similar fashion.
The idea is to add ``horizontal'' arcs in connection models; we call these {\em{info arcs}}. See \cref{fig:bush} for an example bush. An info arc connects two nodes at the same level; thus, these nodes are necessarily not related by the ancestor order in the tree. Info arcs are also labeled by a bounded number of labels, and we assume that every node is connected to itself by an info arc, called an \emph{info loop}. It is still the case that the vertices of $G$ are the leaves of a bush $B$ representing it, but we replace the mechanism of encoding the graph in a bush as follows:\smallskip\\
\noindent\emph{For two vertices $u,v\in V(G)$, whether $u$ and $v$ are adjacent in $G$ depends only on the labels of $u$ and $v$ in $B$, and the label of the lowest info arc in $B$ that connects an ancestor of $u$ with an ancestor of $v$.}\smallskip\\
Note that if there are no info arcs connecting different nodes, then bushes just degenerate to connection models that describe graphs of bounded shrubdepth. However, allowing horizontal passage of information through info arcs allows us to represent much more complicated graphs. This idea is loosely inspired by the mechanics of the parameter {\em{twin-width}} and {\em{contraction sequences}}~\cite{tww1}; see also the work by Bonnet et al.~\cite{tww-perm} for a presentation of twin-width that makes this view more apparent.

We prove that graph classes of structurally bounded expansion are exactly those that decompose into low-depth sparse~bushes.

\begin{theorem}\label{thm:bush-be}
 A class $\DD$ of graphs has 
 structurally bounded expansion if and only if the following condition holds.
  There are $d,\ell\in \N$ such that
  every $G\in \DD$ has a bush $B_G$ representing $G$, where:
  \begin{itemize}
      \item each bush $B_G$ 
      has depth at most $d$ and uses at most $\ell$ labels, and
      \item the class of Gaifman graphs\footnote{The Gaifman graph of a bush $B$ is the graph whose edges are the info arcs of $B$ and the parent-child edges in $B$.} of bushes $\setof{B_G}{G\in \DD}$ has bounded expansion.
  \end{itemize}   
  Moreover, it suffices to use one label for the leaves and two labels for the info arcs.
\end{theorem}

The advantage of \cref{thm:bush-be} over the low-shrubdepth covers from~\cite{GajarskyKNMPST20} is that it provides a single, global decomposition of the graph, rather than a set of decompositions of its local pieces. The advantage of \cref{thm:bush-be} over the lacon and shrub decompositions from~\cite{Dreier21} is that the obtained decomposition is hierarchical\,---\,it has the shape of a bounded-depth tree that represents nested partitions of the vertex set, which is not the case in lacon and shrub decompositions.

We furthermore prove that the class $\setof{B_G}{G\in \DD}$ of bushes 
obtained in \cref{thm:bush-be} can be transduced from the class $\DD$ (see \cref{thm:trandsuction-thm}). This implies characterizations similar to \cref{thm:bush-be} for other properties that are more restrictive than structurally bounded expansion, such as structurally bounded treewidth (see \cref{cor:bushes}).
\begin{figure}[h]\centering
  \includegraphics[page=2,width=0.7\textwidth]{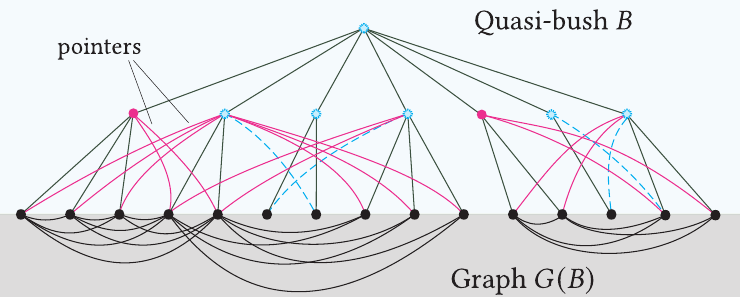}
  \caption{A quasi-bush $B$ and the graph $G(B)$ defined by it. It is a tree equipped with \emph{pointers}, which may connect leaves with internal nodes.
  In this example, the leaves use one label only (which always suffices), and the pointers use two labels (red/blue). Every leaf has a pointer to each of its ancestors $w$,
  omitted in the picture for clarity,
  whose label is indicated by the color of~$w$.
  The vertices of $G(B)$ are the leaves of $B$. Two vertices $u,v$ are adjacent in $G(B)$ if and only if for the lowest ancestor $v'$ of $v$ that is pointed to from $u$,
  the pointer $uv'$ is~red.
  }
  \label{fig:quasi-bush}
\end{figure}

\medskip
The proof of \cref{thm:bush-be} is inspired by the derivation of lacon and shrub decompositions from~\cite{Dreier21}. In particular, it relies on a local variant of the Feferman-Vaught Theorem~\cite{Dreier21,numberOfTypes} rather than on quantifier elimination. The benefit is that this tool works also in the nowhere dense setting. 
Although we were unable to determine whether the analogue of \cref{thm:bush-be} holds 
for structurally nowhere dense classes, we do prove a variant
of the rightwards implication in the theorem.
Namely, we introduce {\em{quasi-bushes}} (see \cref{def:quasi-bush}) and prove that they can be used to decompose graphs from structurally nowhere dense classes (see \cref{fig:quasi-bush}).

\begin{theorem}\label{thm:bush-nd}
 Let $\DD$ be a structurally nowhere dense class of graphs. Then there are $d,\ell\in \N$ such that for every $G\in \DD$, there is a quasi-bush $B_G$ representing $G$, where:
 \begin{itemize}
  \item  each quasi-bush $B_G$ has depth at most $d$ and uses  at most $\ell$ labels, and
  \item the class comprising Gaifman graphs of quasi-bushes $\{B_G\colon G\in \DD\}$ is almost nowhere dense.
 \end{itemize}
 Moreover, it suffices to use one label for the leaves and two labels for the pointers.
\end{theorem}

Here, {\em{almost nowhere denseness}} is a relaxation of nowhere denseness that assumes the preservation of the bound of $\Oh_{\DD,d,\eps}(n^\eps)$ on the weak $d$-coloring numbers. The two notions coincide in the case of hereditary classes (that is, closed under taking induced subgraphs), but, in general, they can differ. For this reason, Theorem~\ref{thm:bush-nd} does not provide a characterization, only an implication in one direction.

The proof of Theorem~\ref{thm:bush-nd} is not merely a simple lift of the proof of Theorem~\ref{thm:bush-be}. The main idea is to replace the parts that exploit weak coloring numbers by a reasoning roughly based on the {\em{Splitter game}}~\cite{GroheKS2017}. In this way, we avoid having factors of the form $\Oh(n^\eps)$ in the bounds on the depth and the sizes of label sets; instead, they are bounded by a constant (depending on $\DD$). 
The proof of \cref{thm:bush-nd} additionally uses an auxiliary result about logical closures in monadically NIP classes (see \cref{thm:NIPLoewenheimSkolem}).
This result may be of independent~interest.

The constant bound on the depth and number of labels in quasi-bushes is crucial for the application, which answers the question of~\cite{GajarskyKNMPST20,BrianskiMPS21} about the existence of small low-shrubdepth covers in structurally nowhere dense~classes. 

\begin{theorem}\label{thm:lsd-nd}
Let $\DD$ be a structurally nowhere dense class of graphs and $p\in \N$ and $\eps>0$ be fixed. Then for every graph $G\in \DD$ one can find a family $\Ff(G)$ of vertex subsets of $G$ with $|\Ff(G)|\leq \Oh_{\DD,p,\eps}(|G|^\eps)$ satisfying the following:
 \begin{itemize}
  \item for every $X\subseteq V(G)$ with $|X|\leq p$, there is $A\in \Ff(G)$ such that $X\subseteq A$; and
  \item the class $\{G[A]\colon G\in \DD, A\in \Ff(G)\}$ has bounded shrubdepth.
 \end{itemize}
\end{theorem}
\cref{thm:lsd-nd} follows easily from \cref{thm:bush-nd}, using an argument similar to that used in~\cite[Section~4]{GajarskyKNMPST20} 
(see \cref{sec:lsd}).

\medskip

Finally, let us remark that, similarly to~\cite{Dreier21,GajarskyKNMPST20}, our results are non-algorithmic. We prove the existence of certain decompositions, but we do not know how to compute them efficiently given only the graph in question. More precisely, the existence of a decomposition is derived from the assumption that the given graph $G$ can be transduced, using a fixed transduction $\trans T$, from a graph $H$ drawn from a fixed sparse class $\CC$. In all cases, the decomposition could be computed efficiently if $H$ was also given on input, but we do not know how to compute it given only $G$. Therefore, our results have no immediate consequences for the complexity of the $\FO$ model-checking problem.

\paragraph*{Acknowledgements} The third author thanks Benedikt Brütsch for very useful technical discussions during the finalization of this manuscript. The fourth author thanks Marcin Briański, Piotr Micek, and Michał T. Seweryn for preliminary discussions on the existence of small low-shrubdepth covers in structurally nowhere dense classes. All authors thank Pierre Ohlmann for proofreading parts of this manuscript.

%!TEX root = main.tex
\section{Preliminaries}\label{sec:prelims}

\paragraph{Graphs}

Unless stated otherwise, the graphs considered in this paper are finite, simple, vertex-colored and undirected. 
This means, a graph $G$ consists of a vertex set $V(G)$, an undirected edge set $E(G)$, and an unspecified number of color classes $C_1,C_2 \dots \subset V(G)$.
Note that the color classes are not necessarily disjoint, so each vertex may have multiple colors or even no color.
If a graph $H$ is an (induced) subgraph of a graph $G$, we require in particular that all vertices $v \in V(H)$ have the same colors in $H$ as in $G$.
A graph $\hat G$ is a \emph{monadic lift} of a graph $G$ if $\hat G$ consists of the vertex set, the edge set, and the color classes of $G$, as well as a number of additional color classes.
%The subgraph of $G$ induced by a set $A \subseteq V(G)$ is denoted by $G[A]$. % and we set $G - A \coloneqq G[V(G) \setminus A]$.
All other common notation for uncolored graphs generalizes to our vertex-colored graphs as expected.

We will also sometimes consider \emph{directed graphs}, equipped with any set $E\subset V\times V$ of \emph{arcs}, where $V$ is the vertex set. A graph can be viewed as a directed graph, by replacing each edge with the pair of arcs in both directions. However, our directed graphs will usually not be colored. 

By a \emph{graph class}, we mean a set of graphs. We assume that every graph class has a fixed finite signature.
This means that, for every graph class, there is a finite palette of colors such that all graphs in the class use only colors from this palette.

To avoid confusion, when we construct treelike decompositions of a graph $G$,
we will use the term \emph{vertex} to refer to elements in $G$
and the term \emph{node} to refer to elements in the treelike decomposition.
Many trees considered in this paper will be rooted. The \emph{depth} of a node $v$ in a rooted tree $T$ is the length (number of edges) of the root-to-$v$ path in $T$, and the \emph{depth} of $T$ is the maximum depth of any node in $T$.

Next, we introduce standard tools for working with graph classes of bounded expansion and nowhere dense graph classes. While the original definitions of these two notions were stated at the beginning of \cref{sec:intro}, we will not use them directly, and instead we rely on the tools to follow.

\paragraph{Weak Coloring Numbers}

Fix a graph $G$ and a total order $\le$ on its vertex set.
We say a vertex $u \in V(G)$ is \emph{weakly $r$-reachable} from a vertex $v \in V(G)$ with respect to $G$ and $\le$
if, in $G$, there is a path of length at most $r$ from $v$ to $u$ such that $u \leq u'$ holds for all $u'$ on the path.
%If in addition, all internal vertices $u'$ of the path satisfy $v \leq u'$, then $u$ is \emph{strongly $r$-reachable} from $v$.\mipi{Why do we need strong reachability?}
We denote by $\wreach_r^{G,\le}[v]$ the sets of all vertices that are weakly $r$-reachable from $v$ with respect to $G$ and $\leq$.
We drop $G$ and $\le$ from the superscript if they are clear from the context.
The \emph{weak $r$-coloring number} of $G$ is
\begingroup
  \allowdisplaybreaks
  \begin{align*}
      \wcol_r(G) & \coloneqq \min_{\text{order $\le$ on $V(G)$}} \wcol_r(G,\le),
      %\col_r(G) & \coloneqq \min_{\text{order $\le$ on $V(G)$}} \col_r(G,\le),
    %\col_r(G) & \coloneqq \min_{\leq \in \Pi(G)} \col_r(G, \leq),
    \shortintertext{where}
    \wcol_r(G, \leq) & \coloneqq \max_{v \in V(G)} |\wreach_r[v]|.
    %\col_r(G, \leq) & \coloneqq \max_{v \in V(G)} |\reach_r[v]|.
  \end{align*}
\endgroup
%The two kinds of generalized coloring numbers are closely related,
%as for all graphs $G$ and numbers $r \in \N$~\cite{coloringdefinition}, we have
%\begin{equation}\label{eq:scolWcolBound}
%\col_r(G) \le \wcol_r(G) \le \col_r(G)^r.
%\end{equation}
For a fixed graph $G$, different values for $r$ may have different orders $\le$ that minimize $\wcol_r(G,\le)$.
Van den Heuvel and Kierstead~\cite{van2019uniform} showed\footnote{The original statement from~\cite{van2019uniform} considers strong coloring numbers. The one here is adapted using known relations between weak and strong coloring numbers, see~\cite[Proposition~4.8]{sparsity}.} that a universal ordering $\le^*$ can be found that is ``good enough'' for all $r \in \N$.
Precisely, for every graph $G$, there exists an ordering $\le^*$ such that, for all $r \in \N$,
\begin{equation}\label{eq:uniformOrder}
    \wcol_r(G,\le^*) \le (2r+1)^r\cdot\wcol_{2r}(G)^{4r^2}.
\end{equation}

It was first observed by Zhu that generalized coloring numbers can characterize bounded expansion and nowhere dense classes~\cite{zhu2009colouring}.

\begin{theorem}[\cite{zhu2009colouring}]\label{thm:col_BE}
    A graph class $\CC$ has \emph{bounded expansion} if and only if, for every $r \in \N$, there exists a number $c\in \N$
    such that $\wcol_r(G) \le c$ for all $G \in \CC$.
\end{theorem}
\begin{theorem}[\cite{zhu2009colouring}]\label{thm:col_ND}
    A graph class $\CC$ is \emph{nowhere dense} if and only if, for every $r \in \N$ and every $\varepsilon > 0$, it holds that $\wcol_r(|H|) \le \Oh_{\CC,r,\eps}(|H|^\eps)$ for every graphs $G\in \CC$ and every subgraph $H$ of $G$.
\end{theorem}
By no longer bounding the generalized coloring numbers for each subgraph, we obtain the following
generalization of nowhere dense graph classes,
which we will use to bound the sparsity of quasi-bushes.
\begin{definition}[Almost Nowhere Dense]\label{def:almostND}
    A graph class $\CC$ is said to be \emph{almost nowhere dense} if for every $r \in \N$ and $\eps > 0$ we have $\wcol_r(G) \le \Oh_{\CC,r,\eps}(|G|^\eps)$ for all graphs $G\in \CC$.
\end{definition}
Note that hereditary graph classes are nowhere dense if and only if they are almost nowhere dense.
As in~\cite{van2019uniform}, \eqref{eq:uniformOrder}
immediately yields the following characterizations
using universal orderings.

\begin{corollary}\label{cor:universalBE}
    A graph class $\CC$ has bounded expansion if and only if,
    for every $G \in \CC$, there exists an order $\le^*$ on $V(G)$ such that for all $r \in \N$, $\wcol_r(|G|,\le^*) \le \Oh_{\CC,r}(1)$.
\end{corollary}
\begin{corollary}\label{cor:universalND}
    A graph class $\CC$ is nowhere dense if and only if,
    for every subgraph $H$ of a graph $G \in \CC$, there is an order~$\le^*$ on $V(H)$ such that,
    for all $\eps > 0$ and $r \in \N$, 
    $\wcol_r(|H|,\le^*) \le \Oh_{\CC,\eps,r}(|H|^\eps)$.
\end{corollary}
\begin{corollary}\label{cor:universalAlmostND}
    A graph class $\CC$ is almost nowhere dense if and only if 
    for every $G \in \CC$ there exists an order $\le^*$ on $V(G)$ such that
    for all $\varepsilon > 0$ and $r \in \N$, 
    $\wcol_r(|G|,\le^*) \le \Oh_{\CC,\eps,r}(|G|^\varepsilon)$.
\end{corollary}

% The \emph{length} of a path equals its number of edges.
% The \emph{depth} of a node $v$ in a rooted tree $T$ with root $r$ is the length of the shortest path from $r$ to $v$.
% The depth of $T$ is the maximal depth of any node in $T$.

%The \emph{height} of $T$ is the maximum depth of a vertex in $T$.\sandra{caution: height vs depth}

%For graphs $G$, $H$, we use the notation $G \subseteq H$ to indicate that $G$ is a \emph{subgraph} of $H$, i.e.\ that $V(G) \subseteq V(H)$ and $E(G) \subseteq E(H)$ holds.
%The subgraph of $G$ induced by a set $A \subseteq V(G)$ is denoted by $G[A]$ and we set $G - A \coloneqq G[V(G) \setminus A]$.

%We say that two vertices $u, v \in V(G)$ are \emph{$r$-separated} by a set $S \subseteq V(G)$ if every %path from $u$ to $v$ of length at most $r$ contains a vertex from $S$.

%A connected graph without cycles is called a \emph{tree}.
%We refer to the vertices of a tree $T$ as the \emph{nodes} of $T$.

%A \emph{rooted tree} is a tree $T$ with a distinguished \emph{root node} $r \in V(T)$.\sandra{check if we direct edges} 
%The \emph{depth} of a vertex $v$ in $T$ is then defined as the number of edges on the shortest path from $r$ to $v$. The \emph{height} of $T$ is the maximum depth of a vertex in $T$.\sandra{caution: height vs depth}

\paragraph{Structures}

We only consider signatures consisting of binary and unary relation symbols. 
We may explicitly say a structure is \emph{binary} to emphasize this.
We see directed and undirected graphs as structures with a single binary edge relation and multiple unary relations, one for each color class.
The universe of a structure $\struct A$ is denoted by $V(\struct A)$.
The \emph{Gaifman graph} of a structure $\struct A$ is the
graph with vertex set $V(\struct A)$ where two elements $u, v \in V(\struct A)$ are connected by an edge if and only if they appear together in some binary relation of $\struct A$.
We say that a class $\CC$ of binary structures has \emph{bounded expansion}, is \emph{nowhere dense}, etc., if the class of its underlying Gaifman graphs has this property.

%Graphs can be viewed as finite structures with a single binary relation.
%For a structure $\struct A$ and a set $X \subseteq V(\struct A)$ we denote by $\struct A[X]$ the \emph{substructure} of $\struct A$ induced by $X$ in the usual way.
%The \emph{Gaifman graph} of a structure $\struct A$ is the graph with vertex set $V(\struct A)$ where two elements $u, v \in V(\struct A)$ are connected by an edge if and only if they appear together in some tuple in some relation of $\struct A$.

%Formulas of \emph{first-order logic} over a signature $\Sigma$ are constructed from atomic formulas $x = y$ and $R(x_1,\ldots,x_k)$, where $x,y,x_1,\ldots,x_k$ are variables and $R \in \Sigma$ is a relation symbol of some arity $k$, using Boolean connectives and quantifiers $\exists, \forall$ as usual.
%All formulas in this paper are first-order formulas, with the usual semantics.

%Consider a formula $\phi(\tup x)$ with free variables $\tup x$.
%For a valuation $\tup u \in V(\struct A)^{|\tup x|}$ of the free variables, we write $\struct A \models \phi(\tup u)$ to indicate that $\tup u$ satisfies $\phi$ in $\struct A$.
%Moreover, we write $\phi(\struct A)$ to denote the structure that consists of the universe $V(\struct A)$ and the relation of all $k$-tuples that satisfy $\phi$ in $G$.
%If $\CC$ is a class of structures, we denote by $\phi(\CC)$ the class of all structures $\phi(\struct A)$ with $\struct A \in \CC$.

\paragraph{Interpretations}

For a structure $\struct A$ and a formula $\phi(\tup x)$ in the signature of $\struct A$, we define \[\phi(\struct A) = \setof{\tup v \in V(\struct A)^{\tup x}}{\struct A\models \phi(\tup v)}.\]
For $\Omega \subseteq V^{\tup x}$ and a subset $A \subseteq V$, we let $\Omega[A] \subseteq \Omega$
be the set of those tuples in $\Omega$ that contain only elements from $A$.

Interpretations use $\FO$ logic to translate between structures.
Let $\Sigma$ and $\Gamma$ be fixed signatures. A (simple) \emph{interpretation} $\trans I$ 
from $\Sigma$-structures to $\Gamma$-structures
consists of a domain formula $\delta(x)$ and a formula
$\phi_R(x_1,\dots,x_k)$ for each $R \in \Gamma$, where $k$ is the arity of $R$.
For a given input $\Sigma$-structure $\struct A$, the output of $\trans I$ is the $\Gamma$-structure $\trans I(\struct A)$
with universe $U=\delta(\struct A)$
and relations $R^{\trans I(\struct A)} = \phi_R(\struct A)[U]$ for each $R \in \Gamma$ of arity $k$.
If $\CC$ is a class of $\Sigma$-structures, then $\trans I(\CC) \coloneqq \setof{\trans I(\struct A)}{\struct A \in \CC}$.
For every formula $\phi(x,y)$, we write $\trans I_\phi$ for the interpretation 
that creates a directed graph with edge set based on $\phi$ and unchanged domain (domain formula $x=x$).

\paragraph{Transductions}

Transductions translate between structures
by first copying and non-deterministically coloring the input structure and then applying a fixed interpretation.
For a number $k \in \N$ and a structure $\struct A$, we define $k \times \struct A$ to be the structure consisting of $k$ disjoint copies of $\struct A$,
together with a new symmetric binary relation $M$ containing all pairs $(v,v')$, where $v$ and $v'$ originate from the same element of $\struct A$.
A {\em{transduction}} from $\Sigma$-structures to $\Gamma$-structures consists of
\begin{itemize}
    \item a number $k \in \N$,
    \item unary relation symbols $U_1,\dots,U_\ell$, and
    \item an interpretation $\trans I$ from $\Sigma \cup \set{M,U_1,\dots,U_\ell}$ to $\Gamma$.
\end{itemize}
For a transduction $\trans T$ and input $\Sigma$-structure $\struct A$, the output $\trans T(\struct A)$ consists of all $\Gamma$-structures $\struct B$
such that there exists a coloring $\struct{\hat A}$ of $k \times \struct A$ with fresh unary predicates $U_1,\dots,U_\ell$
such that $\struct B= \trans I(\struct{\hat A})$.
If $k=1$, we say the transduction is \emph{non-copying}. Transductions are closed under composition.

A class $\DD$ of graphs has \emph{structurally bounded expansion} if there
exists a class $\CC$ of graphs with bounded expansion and a transduction $\trans
T$ such that $\DD \subseteq \trans T(\CC)$.
Similarly,
a class $\DD$ of graphs is called \emph{structurally nowhere dense} if there
exists a nowhere dense graph class $\CC$ and a transduction $\trans T$ such that
$\DD \subseteq \trans T(\CC)$.
We observe that in both cases we can assume $\trans T$ to be non-copying:

\begin{lemma}\label{lem:ND-nocopy}
    Let $\DD$ be a structurally nowhere dense graph class.
    Then there exists a nowhere dense graph class $\CC$ and a non-copying transduction $\trans T$
    such that $\DD \subseteq \trans T(\CC)$. If $\DD$ is moreover of structurally bounded expansion, then we may choose $\CC$ and $\trans T$ so that $\CC$ has bounded expansion.
\end{lemma}
\begin{proof}
    We prove the claim for nowhere dense classes, the proof for the bounded expansion case is the same.

    By assumption, there is a nowhere dense graph class~$\CC'$ and a transduction $\trans T'$ such that $\DD \subseteq \trans T'(\CC')$. 
    %We first argue that without loss of generality we may assume that $\trans T$ is non-copying. 
    If $\trans T'$ creates $p$ copies of the universe, then we may consider the
    class $\CC$ consisting of lexicographic products of graphs from $\CC'$ with
    $K_{p+1}$. It is well-known that $\CC$ is still nowhere
    dense~\cite[Proposition~4.6]{sparsity}, and it is straightforward to see that $\DD$ can
    now be transduced from $\CC$ by a non-copying transduction~$\trans T$.
    Further, again by~\cite[Proposition~4.6]{sparsity}, if $\CC'$ has bounded expansion, then so does $\CC$.
\end{proof}

In fact,
for classes with structurally bounded expansion, we have a much stronger statement.
For two classes $\CC$ and~$\DD$ and a transduction $\trans T$ on $\CC$, we write $\trans T: \CC\tto \DD$ to indicate that
$\trans T(\CC) \subseteq \DD$.

\begin{theorem}[Proposition 18 of~\cite{GajarskyKNMPST20}]\label{thm:sparsification}
  Let $\DD$ be a class of graphs with structurally bounded expansion.
  Then there are a graph class $\CC$ of bounded expansion,
   a non-copying transduction $\trans T\from \CC\tto \DD$, 
   and a transduction $\trans T'\from \DD\tto \CC$
   such that, for every $H\in \DD$,
   we have $H\in \trans T(\trans T'(H))$.
\end{theorem}

The following folklore lemma is helpful in constructing transductions.
The \emph{star chromatic number} of a graph $G$ is the smallest number of colors needed to vertex-color $G$ such that every vertex receives exactly one color and the subgraph induced by the union of each pair of color classes is a star forest (a disjoint union of stars). Such a coloring is a \emph{star coloring} of $G$.
Every class of bounded expansion has bounded star chromatic number~\cite[Theorem~7.7]{sparsity}.

\begin{lemma}\label{lem:direction transduction}
  Let $\widehat \CC$ be a class of binary structures such that the class $\CC$ of Gaifman graphs of structures in $\widehat \CC$ has bounded star chromatic number. Then there is a transduction $\wh{\trans T}$ such that $\widehat \CC \subseteq \wh{\trans T}(\CC)$.
\end{lemma}
\begin{proof}
    First consider the case when $\CC$ is a class of star forests. Then 
    given a star forest $S\in\CC$ that is the Gaifman graph of $\wh S\in\wh\CC$,
    the transduction $\wh{\trans T}$
    first introduces a unary predicate (color) $Q$,
    with the intention that   
    $Q$ marks the center in each star of $S$ (in stars with two vertices, we designate any vertex as the root).
  Next, by introducing several unary predicates, the transduction labels each node $v\in S$ by the atomic type\footnote{The {\em{atomic type}} of a pair of vertices is the information on which unary and binary predicates are satisfied by vertices in this pair.} of the pair $(v,v')$ in $\wh S$, where
  $v'\in Q$ is the center of the star in which $v$ is contained (possibly $v=v'$). Finally, the
  transduction introduces a relation $R$, for each $R$ in the signature of~$\wh\CC$. 
  Each relation $R$ is described by a formula $\psi_R(x,y)$ expressing the following~assertions:
  \begin{itemize} 
    \item either $x=y$, or $x$ and $y$ are adjacent;
    \item if $x$ is such that $\neg Q(x)$, then the label of $x$ is an atomic type that implies $R(x,y)$;
    \item if $y$ is such that $\neg Q(y)$, then the label of $y$ is an atomic type that implies $R(x,y)$;
    \item if $x=y$ and $Q(x)$, then the label of $x$ is an atomic type that implies $R(x,x)$.
  \end{itemize}
  %The formula $\phi_R(x,y)$ describing $R$
  %holds if the following conditions are satisfied:
  %\begin{itemize}
    %\item $Q(y)$ holds, 
    %\item either $x=y$, or $x$ and $y$ are adjacent,
    %\item the label of $x$ is an atomic type that implies the formula $\phi_R(x,y)$.
  %\end{itemize}
  This finishes the case when $\CC$ is a class of star forests.
  
      Now consider the general case, and suppose that $\CC$ has star chromatic number bounded by some $k$. Then 
      for every $\widehat G\in\widehat\CC$, the Gaifman graph $G$ of $\widehat G$ can be (disjointly) colored using $k$ colors so that for any two colors $C$ and $D$, the induced subgraph  $G[C\cup D]$ is a star forest.
    Let $\widehat G[C\cup D]$ be the substructure of $\widehat G$ induced by $C\cup D$, and let $\widehat {\cal S}$ be the class of all structures of the form $\widehat G[C\cup D]$ that can be obtained as above, in any $G\in \CC$.
    Then the Gaifman graphs of the structures in $\widehat{\cal S}$ 
    are star forests, so we can apply the special case above,
    obtaining a transduction ${\wh{\trans T}_{\cal S}}$
    such that $\widehat G[C\cup D]\in \wh{\trans T}_{\cal S}(G[C\cup D])$ for every $G\in \CC$, star coloring of $G$ using $k$ colors,
    and two color-classes $C,D$ in this star~coloring.
  
    Given a graph $G$, we define a transduction $\trans T$ that  first colors $G$ using $k$ disjoint colors to obtain a star coloring.
    Then for each pair of colors $C,D$, it applies ${\wh{\trans T}_{\cal S}}$
    to $G[C\cup D]$ to obtain $\wh G[C\cup D]$. Then it takes the union of 
    $\wh G[C\cup D]$ over all $k\choose 2$ pairs $C,D$ of colors, yielding $\wh G$.
\end{proof}
 
\paragraph{First-Order Types}

Let $G$ be a graph, $\tup x$ be a tuple of variables, and $\tup v \in V(G)^{\tup x}$.
The \emph{$q$-type} of $\tup v$ in $G$ is
the set $\tp^q_G(\tup v)$ of all $\FO$ formulas $\phi(\tup x)$ of (quantifier) rank at most $q$ with the same signature as $G$
such that $G \models \phi(\tup v)$.
We omit the subscript $G$ if the graph is clear from the context.
We assume all formulas to be normalized so that $q$-types are finite.
In particular, their size can be bounded by a function of $|\tup x|$, $q$, and the signature~of~$G$. For a vertex $u$ and tuple $\tup v$, by $u\tup v$ we mean the tuple obtained from $\tup v$ by mapping a fresh variable to $u$.

We say that two vertices $u, v \in V(G)$ are \emph{$r$-separated} in a graph $G$ by a set $S \subseteq V(G)$ if every path from $u$ to $v$ of length at most $r$ contains a vertex from $S$.
Based on this notion of separation, we use the following local composition result inspired by the Feferman--Vaught theorem~\cite{FV59}.

\begin{lemma}[{\cite[Lemma 15]{numberOfTypes}, \cite[Theorem 4]{Dreier21}}]\label{lem:separation and types}
  For every formula $\phi(x,y)$, there are numbers $r,q\in\N$
  such that for every graph $G$, tuple of variables $\tup x$, tuple $\tup s \in V(G)^{\tup x}$,
  and all $u,v\in V(G)$ that are $r$-separated by $\tup s$ in $G$,
  the truth value of $\phi(u,v)$ is determined by the (ordered) pair of types $\tp^{q}(u\tup s)$ and $\tp^{q}(v\tup s)$. More precisely, there is a binary relation 
  $R$ on the set $\setof{\tp^q(v\tup s)}{v\in V(G)}$ 
  such that for any 
  two vertices $u,v\in V(G)$ that are $r$-separated by $\tup s$ in $G$,
  $\phi(u,v)$ holds in $G$ if and only if 
  the pair 
  $(\tp^{q}(u\tup s),\tp^{q}(v\tup s))$ belongs~to~$R$.
\end{lemma}

%!TEX root = main.tex
\section{Bushes}\label{sec:bushes}%

In this section, we develop a succinct representation for graphs from a fixed class with structurally bounded expansion. The crucial decomposition notion is described below.

\begin{definition}
  A \emph{bush} $B$ consists of:
  \begin{itemize}
    \item a rooted tree $T$ in which all leaves have equal depth. The leaf set of $T$ is denoted by $\leaves(B)$,
    and the set of nodes of $T$ is denoted by $V(B)$. The \emph{depth} of $B$ is the depth of~$T$;
    \item a symmetric, reflexive binary relation $I\subset V(B)\times V(B)$, whose elements are called \emph{info arcs}, and are such that both endpoints have equal depth in $T$. This depth is the \emph{depth} of the info arc;
    \item a labeling function $\lambda\from \leaves(B)\to \Lambda$, where $\Lambda$ is a finite set of \emph{labels};
    \item a labeling function $\lambda^I\from I\to 2^{\Lambda\times \Lambda}$.
  \end{itemize}

  Let $u,v \in \leaves(B)$. The \emph{lowest info arc}
  above $(u,v)$ is the info arc $(u',v')\in I$ with largest depth such that $u'$ is an ancestor of $u$ and $v'$ is an ancestor of~$v$. 

  Every bush $B$ defines a directed graph $G(B)$ whose vertices are the leaves of $B$ and in which there is an arc $(u,v)$ between distinct leaves $u$ and $v$ if and only if $(\lambda(u),\lambda(v))$ belongs to the label $\lambda^I(u',v')$ of the lowest info arc $(u',v')$ above $(u,v)$. We say that $B$ \emph{represents} $G(B)$.
\end{definition}

Note that, a priori, bushes represent directed graphs. 
If we want to represent an undirected graph $G$
using a bush~$B$, then we formally require that $G(B)$
is the directed graph corresponding to $G$ where each edge in $G$ is replaced by two oppositely-oriented arcs in $G(B)$.  Further, note that the info arc relation is symmetric, but info arcs are directed. That is, for two nodes $a,b$ related in $I$, the info arcs $(a,b)$ and $(b,a)$ may receive different labels under $\lambda^I$.

We represent bushes as relational structures in the natural way, using one binary parent relation and $|2^{\Lambda\times \Lambda}|$ binary relations for info arcs. So we may speak about Gaifman graphs of bushes and of classes of bushes of bounded expansion.

% The \emph{Gaifman graph} of a bush $B$ 
% is the graph whose vertices are the nodes of $B$,
% and whose edges connect two vertices that are either related by the parent-child relation in $B$, or are related by an info arc. We say a class $\BB$ of bushes has \emph{bounded expansion} if the class of its Gaifman graphs has bounded expansion.

Now we can restate \cref{thm:bush-be} as follows.

\begin{theorem}\label{thm:decomposition trees for be}
 A class $\DD$ of graphs has structurally bounded expansion if and only if there is a class $\BB$ of bushes of bounded depth, bounded expansion, and using a fixed finite set of labels~$\Lambda$, such that for every $H\in\DD$, there is some $B\in \BB$ that represents $H$.
 Moreover, we can take $|\Lambda|=1$.
\end{theorem}

The `moreover' part says that we may assume that the bushes have all leaves labeled with the same label $\bullet$, and there are two 
possible labels on the info arcs: the empty and the full binary relation on $\set\bullet$.
The mechanism 
of defining adjacency works by creating a directed edge $(u,v)$ if the lowest info arc 
above $u$ and $v$ is labeled with $\set{(\bullet, \bullet)}$. Moreover,
it is easy to see that 
 in such bushes defining undirected graphs, the info arcs may be assumed to be undirected:
if there is an info arc $(a,b)$, then the info arc $(b,a)$ has the same label as $(a,b)$, as is depicted in \cref{fig:bush}.

Thus, bushes precisely characterize graph classes of structurally bounded expansion. 
We first prove the right-to-left implication in \cref{thm:decomposition trees for be}.
Let $\BB_0$ be the class of Gaifman graphs of bushes in $\BB$. Then $\BB_0$ has bounded expansion.
Note that bushes from $\BB$ can be represented as binary structures over a fixed signature, consisting of the binary parent relation, one unary relation for each label $a \in \Lambda$, and one binary relation for each binary relation $R\subset {\Lambda\times \Lambda}$.
Since classes of graphs of bounded expansion have bounded star coloring number, by \cref{lem:direction transduction} there is a transduction 
$\trans T$ such that $\BB\subset \trans T(\BB_0)$.
It is easy to see that there is an interpretation $\trans I$ that on input a bush $B\in \BB$ outputs the directed graph $G(B)$: the interpretation $\trans I$ restricts the universe to the leaves of $B$ and creates an arc between two leaves $u,v$ as indicated by the lowest info arc above $u$ and $v$. This info arc can be determined by an $\FO$ formula, 
since $B$ has bounded depth and uses only labels from $\Lambda$.
Hence, $\DD\subset \trans I(\trans T(\BB_0))$.
In particular, $\DD$ has structurally bounded expansion,
as required.

We therefore focus on the proof of the left-to-right implication in the theorem,
as well as reducing the number of labels to $1$.
Before doing that, we make some preparatory assumptions
concerning $\DD$. 

\subsection{Preparation}
Say that a graph class $\DD$ is {\em{represented}} by a class 
of bushes $\BB$ if every $H\in\DD$ is represented by some bush $B_H\in \BB$.
The  left-to-right implication
in \cref{thm:decomposition trees for be}
is restated below.

\begin{proposition}\label{prop:imp}
  Let $\DD$ be a graph class with structurally bounded expansion.
  Then $\DD$ is represented by some class $\BB$ of bushes with bounded expansion,
  of bounded depth, and using a bounded number of labels.
  Moreover, one label suffices.
\end{proposition}

Let $\DD$ be a graph class of structurally bounded expansion.
It follows from \cref{lem:ND-nocopy} that there is a class $\CC$ of (colored) graphs that has bounded expansion and a formula $\phi(x,y)$ 
such that $\phi(x,y)$ implies $\phi(y,x)\land (x\neq y)$
(so that $\trans I_\phi$ produces graphs)
and $\DD$ is contained in the hereditary closure of $\trans I_\phi(\CC)$ (that is, the closure under taking induced subgraphs).
The following simple lemma implies that, to prove \cref{prop:imp},
we may assume that $\DD=\trans I_\phi(\CC)$.
%In the proof, we note that for every bush $B$ and every set of leaves $W$, 
%the subgraph of $G(B)$ induced by $W$ is represented by the bush obtained by restricting $B$ to nodes of $W$ and their ancestors.

\begin{lemma}\label{lem:restrict}
  Let $\DD$ be a class of graphs and let $\DD'$ be 
  its hereditary closure.
  If $\DD$ 
is represented by some class of bushes $\BB$ with bounded expansion,  depth $d$, and  labels $\Lambda$,
 then also $\DD'$ is represented by some class $\BB'$ of bushes with bounded expansion,  depth $d$, and labels $\Lambda$.
\end{lemma}
\begin{proof}
    For a bush $B$ and a subset of its leaves $W$,
    let $B[W]$ be the bush obtained from $B$ by taking only the leaves in $W$, their ancestors,
    as well as the tree edges and info arcs connecting them.
    Observe that if $H$ is the graph represented by $B$,
    then the subgraph $H[W]$ is represented by $B[W]$.
    Further, the class of all subgraphs of graphs from a class with bounded expansion still has bounded expansion.
    Hence, the class of~bushes 
    \[\BB'=\setof{B[W]}{B\in \BB, W\subseteq \leaves(B)}\]
    has bounded depth, uses a bounded number of labels,
    has bounded expansion, and now for each $H\in \DD$ and $W\subset V(H)$,
    the graph $H[W]\in \DD'$ is represented by some bush in this class.
    Since every graph in $\DD'$ is of the form $H[W]$ for some $H\in\DD$ and $W\subset V(H)$, this yields the conclusion.
\end{proof}

To prove \cref{prop:imp}, we therefore need to prove that, 
for every class $\CC$ of (colored) graphs with bounded expansion and for every formula $\phi(x,y)$,
the class $\trans I_\phi(\CC)$ 
is represented by a class $\BB$ of bushes of bounded expansion, bounded depth, and with a bounded number of labels.
Moreover, we need to show that we can reduce the number of labels to $1$. For the remainder of the entire \cref{sec:bushes}, let us fix the following objects:
\begin{itemize}
  \item a class $\CC$ of (colored) graphs with bounded expansion, 
  \item a formula $\phi(x,y)$ in the signature of $\CC$, and
\item numbers $q$ and $r$ provided by \cref{lem:separation and types} applied to~$\phi$.
\end{itemize}
With these fixed, \cref{prop:imp} follows directly by combining the following lemmas.

% 
% \medskip
% The following objects are fixed for the rest of \cref{sec:bushes}:
% \begin{itemize}
%   \item a class $\CC$ of colored graphs with bounded expansion, \item a formula $\phi(x,y)$ in the signature of $\CC$,
% \item numbers $q$ and $r$, as given by \cref{lem:separation and types} applied to~$\phi$,
% \item a number $d$ that will later be specified in the proof of \Cref{lem:beBushes}.
% \skin{I find this quite ugly, to first fix it but leave unknown, then release it for lemma 7, then fix it for lemma 8. It might be better to release it fully, state Lemma 6 with "Let $d$ ..." and state Lemma 8 with "Let $d$ be as in Lemma 7", and fix d after lemma 8.}
% \end{itemize}
% Our goal is to construct for every $G\in \CC$ a bush representing $\trans I_\phi(G)$, so that the obtained class of bushes has bounded expansion and depth,
% and, furthermore, the bushes use one label only.
% We achieve this by proving the following lemmas.

\begin{lemma}\label{lem:bush construction}
  For each $G\in\CC$ and order $\le$ on $V(G)$, when letting $d \coloneqq \wcol_r(G,\le)$,
  there is a bush $B(G,\le)$ that represents $\trans I_\phi(G)$,
  has depth $d$, and uses a label set depending only on $q$, $r$, and~$d$.
\end{lemma}

\begin{lemma}\label{lem:beBushes}
  There exist $d\in \N$ and a class $\cal B$
  of bushes with bounded expansion
  such that,
  for every $G \in \CC$, there is an order $\le$ on $V(G)$ such that $\wcol_r(G,\le) \le d$ and  $B(G,\le) \in \cal B$.
\end{lemma}

\begin{lemma}\label{lem:one label}
  Let $d \in \N$, let $B$ be a bush of depth $d$, and let $\Lambda$ be the label set of $B$. Then there is a bush $B'$ representing the same graph such that $B'$ has depth $d+1$, uses only a single label, and satisfies 
  $\wcol_s(B') \le |\Lambda| \cdot \wcol_s(B) + 1$ for all $s \in \N$.
\end{lemma}
In the above, the coloring numbers refer to the underlying Gaifman graphs of bushes.

%\medskip
%By the argument presented above, the three lemmas together yield \cref{prop:imp}, which in turn is sufficient to finish the proof of \cref{thm:decomposition trees for be}.
\cref{lem:bush construction} and \cref{lem:beBushes} are proved in Sections \ref{sec:bush-construction}, \ref{sec:bushes-be},  respectively. Note that \cref{lem:one label} is only used to reduce the number of labels to $1$.
Finally, in \cref{sec:transduction section}, we state \cref{thm:trandsuction-thm}, which says that the bush $B(G,\le)$ can be produced by a transduction that takes $G$ on input. We also present some interesting implications of this statement. The proof of \cref{thm:trandsuction-thm} is in the appendix.

%For a number $r\in\N$, a graph $G$ and $\le \in \Pi(G)$, let $\wreach_r[v]$ denote the set of vertices of $G$ that are weakly $r$-reachable from $G$ with respect to $\le$.

\paragraph{Weak reachability.}
Before proceeding with the proofs, we first collect two useful insights about weak reachability sets. Both of them are standard, but fundamental.

\begin{lemma}\label{lem:weak-separation}
  Let $G$ be a graph, $\le$ be an order on $V(G)$, and $r\in \N$.
  Then any two vertices $u$ and $v$ of $G$ are $r$-separated in $G$ by the set
  $\wreach_r[u]\cap \wreach_r[v]$.
\end{lemma}
\begin{proof}
 Let $P$ be any path of length at most $r$ connecting $u$ and $v$ and let $w$ be the $\leq$-minimum vertex on $P$. Then the subpath of $P$ from $u$ to $w$ witnesses that $w\in \wreach_r[u]$, while the subpath from $w$ to $v$ witnesses that $w\in \wreach_r[v]$. Consequently, $w\in \wreach_r[u]\cap \wreach_r[v]$.
\end{proof}

\begin{lemma}\label{lem:ordering wreach}
  Let $G$ be a graph, $\le$ be an order on $V(G)$, and $r\in \N$.
  Then for every $u\in V(G)$ and $v,v'\in \wreach_r[u]$, it holds that
  $v\le v'$ if and only if $v\in\wreach_{2r}[v']$.
\end{lemma}
\begin{proof}
The ``if''-part holds by definition, since no element in $\wreach_{2r}[v']$ is larger than~$v'$. For the ``only if''-part, it suffices to note that if $P$ and $P'$ are paths witnessing that $v,v'\in \wreach_r[i]$, respectively, then the concatenation of $P$ and $P'$ is a walk that witnesses that $v \in \wreach_{2r}[v']$.
\end{proof}

% \subsection{Normal form of transductions}
% We will use the following statement, 
% providing a normal form of transductions.

% For a formula $\phi(x,y)$, let $\trans I_\phi$ be the interpretation
% that does not change the domain and creates for an input structure $\struct A$ only a single binary relation $\setof{(v,w) \in V(\struct A)}{\struct A \models \phi(v,w)}$.
% Then for a structure $\struct A$ or class $\CC$ of structures, we write
% $\phi(\struct A)$ and $\phi(\CC)$ as a synonym for $\trans I_\phi(\struct A)$ and $\trans I_\phi(\CC)$.

% \begin{lemma}\label{lem:trans-normal}
%   The following conditions are equivalent for a class $\DD$  of graphs:
%   \begin{itemize}
%     \item $\DD$ has structurally bounded expansion, 
%     that is, $\DD\subset \trans T(\CC)$ for some class $\CC$ with bounded expansion and transduction $\trans T$,
%     \item there is a class $\CC$ of colored graphs that has bounded expansion and a formula $\phi(x,y)$ in the signature of $\CC$, such that $\DD$ is contained in the hereditary closure of $\trans I_\phi(\CC)$.
%   \end{itemize}
% \end{lemma}

\subsection{Construction of bushes}\label{sec:bush-construction}
We prove \cref{lem:bush construction} by constructing a bush $B(G,\le)$ representing each graph 
of the form $\trans I_\phi(G)$, for $G\in \CC$.
The bush is parameterized by an order $\le$ on $V(G)$.
It will be clear from the construction that $B(G,\le)$ has depth $d = \wcol_r(G,\le)$ and uses a number of colors bounded in terms of $q$, $r$, and $d$. In \cref{sec:bushes-be}, we will show that the obtained class of bushes has bounded expansion, that is, we will prove \cref{lem:beBushes}.

Fix $G\in \CC$ and an order $\le$ on $V(G)$ 
with ${\wcol_r(G,\le)}\leq d$. We construct $B(G,\le)$ as~follows.
  \paragraph{Tree.}
For $0\le i\le d$ and $v \in V(G)$, let $\first_i(v)$ 
be the sequence of length $i$ consisting of 
the $i$ smallest (with respect to $\le$) elements of  $\wreach_r[v]$, in increasing order. In case $|\wreach_r[v]|<i$, we pad the sequence by repeating $v$ so that it has length exactly~$i$.
Define a tree $T$ whose nodes at depth $i$, where $0\le i\le d$, are the sequences $\first_i(v)$ for all $v\in V(G)$.
For two nodes $X, Y$ of $T$, declare $X$ an ancestor of $Y$ in $T$ if $X$ is a prefix of $Y$. Thus, the root of $T$ is the empty sequence and all leaves of $T$ are at depth~$d$.

\paragraph{Leaf labels.}
Note that the mapping $\first_d\from V(G)\to V(T)$ maps $V(G)$ bijectively to the leaves of $T$. 
We therefore identify the leaves of $T$ with $V(G)$ via this mapping.

% Fix $v\in V(G)$ and let $S \coloneqq \wreach_r[v]$. Then $|S|\le d$ and
% the vertices of $S$ are distinctly colored by the restriction of $c\from V(G)\to C$ to $S$.
%Let $G'$ be the graph obtained from $G$ by forgetting all color classes that do not occur in $\phi(x,y)$.
%Then the signature of $G'$ has size at most $|\phi|$, while also $\trans I_\phi(G)=\trans I_\phi(G')$.
Label the leaf $v$ of $T$ by the type $\tp^q(v\,\first_d(v))$ of the tuple $\first_d(v)$ with $v$ prepended to it.
This gives a label $\lambda(v)$ from a finite set $\Lambda$ of labels that depends only on $q$, $d$, and~$\CC$, but is independent of the choice of $G\in \CC$ and~$v\in V(G)$.

\paragraph{Info arcs.}
In what follows, whenever $X$ is a node of $T$, 
that is, a sequence of vertices of $G$ of the form $\first_i(v)$,
we also treat $X$ as the underlying set of vertices,
when using set-theoretic notation such as $v\in X$ or $X\subset Y$.
We will also use the terms $\max X$ and $\min X$, respectively, to refer to the maximum and minimum elements in $X$ with respect to $\le$.
Note that $\max X$ equals the last element of $X$, since $X$ is non-decreasing. 
We define the info arcs to be the set $I\subset V(T)\times V(T)$ consisting of all pairs $(X,Y)$ of non-root nodes at the
same depth in $T$ satisfying the following condition:
\[\max X\in Y\qquad\textrm{or}\qquad \max Y\in X.\]
We additionally add the info loop $(\emptyset,\emptyset)$.
Note that, thus, each node of $T$ is equipped with an info loop.

\begin{lemma}\label{lem:lia}
  Fix an info arc $(X,Y)\in I$. 
  Then for all $u,v\in V(G)$ such that $(X,Y)$ is the lowest info arc above $(u,v)$, the truth value of $\phi(u,v)$ in $G$ depends only on the label of $u$ and the label of $v$.
  More precisely, there is a binary relation $R\subset \Lambda\times \Lambda$ such that  $G\models \phi(u,v)$  if and only if $(\lambda(u),\lambda(v))\in R$, for all $u,v$ as above.
\end{lemma}
\begin{proof}
  Let $0\le i\le d$ be the depth of the info arc $(X,Y)$.
Let $u$ and $v$ be vertices of $G$, viewed as leaves of $T$,
such that  $(X,Y)$ is the lowest info arc above $(u,v)$.
In particular, $X$ is an ancestor of $u$ and $Y$ is an ancestor of $v$,
and $X=\first_i(u)$ and $Y=\first_i(v)$.
We show that following inclusion holds:
\begin{align}\label{eq:first-separation}
  \wreach_r[u]\cap \wreach_r[v]\subset
  \first_i(u)\cap \first_i(v).    
\end{align}
Let $w$ be any element of $\wreach_r[u]\cap \wreach_r[v]$,
and assume that $w$ is the $j$th smallest element of $\wreach_r[u]$ and the $k$th smallest element of $\wreach_r[v]$. Suppose by symmetry that $k\le j$. Then $w=\max(\first_j(u))$ and $w\in\first_k(v)\subset \first_j(v)$,
so in particular, there is an info arc between $\first_j(u)$ and $\first_j(v)$. Hence $j\le i$, as $i$ is the depth of the lowest info arc above $(u,v)$. This proves~\eqref{eq:first-separation}. 

By~\eqref{eq:first-separation} and Lemma~\ref{lem:weak-separation},
the set $S \coloneqq X\cap Y\supseteq\wreach_r[u]\cap \wreach_r[v]$ is an $r$-separator between $u$ and $v$. Let $\tup s$ be the tuple enumerating $S$ in increasing order with respect to $\le$.
Let $R_S$ be the binary relation on $\setof{\tp^q(v\tup s)}{v\in V(G)}$ given by 
Lemma~\ref{lem:separation and types}.

To prove the lemma, it suffices to show that, if $u$ and $u'$ are two descendants of $X$ with the same label $\lambda(u)=\lambda(u')$, and $v$ and $v'$ are two descendants of $Y$ with the same label $\lambda(v)=\lambda(v')$, and $(X,Y)$ is the lowest info arc above $(u,v)$, as well as the lowest info arc above $(u',v')$,
then $G\models \phi(u,v)\Leftrightarrow \phi(u',v')$. Indeed, then $R$ can be defined as the set of all pairs $(\lambda_1,\lambda_2)\in \Lambda\times \Lambda$ such that there are $u,v$ as above with $(\lambda_1,\lambda_2)=(\lambda(u),\lambda(v))$ and $G\models \phi(u,v)$.

First we argue that 
\begin{align}\label{eq:types}
\tp^q(u\tup s)=\tp^q(u'\tup s).  
\end{align}
This is because $S\subset X = \first_i(u) = \first_i(u')$. Hence, 
an element $s\in S$ occurs in the $j$th position of $\first_i(u)$ if and only if it occurs in the $j$th position of $\first_i(u')$.
Now \eqref{eq:types} follows from the fact that $u$ and $u'$ have equal labels.

An equality analogous  to~\eqref{eq:types} holds for $v$ and $v'$.
By the definition of $R_S$,
we have that $G\models \phi(u,v)$ if and only if 
$(\tp^q(u\tup s),\tp^q(v\tup s))\in R_S$.
By the equality~\eqref{eq:types} and its version for $v$ and $v'$,
this is equivalent to $(\tp^q(u'\tup s),\tp^q(v'\tup s))\in R_S$,
which in turn is equivalent to $G\models \phi(u',v')$, as required.
\end{proof}

We label the info arc $(X,Y)$ by the relation $R$ 
given by the lemma. Note that this relation might not be symmetric, and that 
$(Y,X)$ might be labeled by a different relation.

This finishes the construction of the bush $B(G,\le)$.
By \cref{lem:lia}, $\phi(u,v)$ holds if and only if the label of $u$ and the label of $v$ form a pair belonging to the relation $R$ that labels $(X,Y)$, where $(X,Y)$ is the lowest info arc above $(u,v)$. Hence,  the constructed bush indeed represents $\trans I_\phi(G)$.
This proves \cref{lem:bush construction}.

\subsection{Bounding the coloring numbers of bushes}
\label{sec:bushes-be}
We now prove \cref{lem:beBushes}. That is, for every $G\in\CC$, we need to pick an order $\le_G$ so that $\wcol_r(G,\le_G)$ is uniformly bounded and the resulting class $\BB$ of bushes of the form $B(G,\le_G)$ as described in \cref{sec:bush-construction} has bounded expansion. % CHECK: added the requirement about the weak coloring number
    
By \cref{cor:universalBE}, for every $s\in \N$, there exists $d_s\in \N$ such that, for every $G\in\CC$, there is 
an order $\le_G$ on $V(G)$ satisfying for all $s' \in \N$ that $\wcol_{s'}(G,\le_G)\leq d_{s'}$.
As the number $d$ whose existence is postulated in \cref{lem:beBushes}, we choose $d\coloneqq d_r$.

 Fix $G\in\CC$, 
the order $\le$ on $V(G)$ equal to $\le_G$, 
 and let $B \coloneqq B(G,\le)$ be the bush constructed in the proof of \cref{lem:bush construction} (\cref{sec:bush-construction}), with underlying tree $T$.
By construction, $B$ has depth bounded by $d$. 
We extend the order $\le$ from $V(G)$ to all nodes of $B$ so that the root of $B$ comes first and for any two non-root nodes $X,Y$ of $B$ with $\max X \le \max Y$, it holds that $X \le Y$.

We prove the result by showing that for all $s \in \N$, we have:
\begin{equation}\label{colbound_of_B_be}
\mspace{-8mu}\wcol_s(B,\le) \mspace{-1mu}\le\mspace{-1mu} \wcol_{2sr}(G,\le) \mspace{-1mu}\cdot\mspace{-1mu} \wcol_{r}(G,\le) \mspace{-1mu}\cdot\mspace{-1mu} 2^{\wcol_{2r}(G,\le)} \mspace{-1mu}+\mspace{-1mu} 1.
\end{equation}
Note that with \cref{thm:col_BE}, this will prove that $\BB$ has bounded expansion.

%For sets $X \subset V(G)$ let $\max X$ and $\min X$ be the maximum and minimum of $X$ with respect to $\le$.
%We define a function $\rep(X) \from V(T) \to V(G)$
%$$
%\rep(X)=
%\begin{cases}
%    \rep(X) = \max X \quad & \text{$X$ internal node of $T$}, \\
%    \rep(X) = X \quad & \text{$X$ leaf of $T$},
%\end{cases}
%$$
We will need the following three observations.
\begin{claim}
\label{lem:rep_inverse_count_be}
For all $v \in V(G)$, $|\setof{Y \in V(T)}{\max Y=v}| \le \wcol_{r}(G,\le) \cdot  2^{\wcol_{2r}(G,\le)}$.
\end{claim}
\begin{proof*}
Let $Y$ be any node such that $\max Y = v$.  Then it holds that $Y = \first_i(u)$ for some $u \in V(G)$ and $i$. By applying \cref{lem:ordering wreach} to $u$, $v$, and every $v'\in Y$, we conclude that $Y\subseteq \wreach_{2r}^{G,\le}[v]$. There are at most $2^{\wcol_{2r}(G,\le)}$ subsets of $\wreach_{2r}^{G,\le}[v]$ and since there are at most $\wcol_r(G,\le)$ nodes with the same underlying set as $Y$ (due to padding in the definition of $\first_i$), the result follows.
\end{proof*}

\begin{claim}
\label{lem:path_be}
Let $X,Y$ be two non-root nodes of $B$ that are adjacent, either by the parent relation in $T$, or via an info arc. Then there exists a path $\pi$ of length at most $2r$ in $G$ with endpoints $\max X$ and $\max Y$ such that 
$$\min(\max X, \max Y) \le \min V(\pi).$$
\end{claim}
\begin{proof*}
We prove that in each case, it holds that $\max X \in \wreach_{2r}[\max Y]$ or $\max Y \in \wreach_{2r}[\max  X]$. Then any path $\pi$ witnessing this weak reachability satisfies the premise of the claim.

Assume $(X,Y)$ is a parent-child edge, meaning that $X$ is a prefix of $Y$. Then $X \subset Y$ and $\max X \le \max Y$. Let $v$ and $i$ be such that $Y = \first_i(v)$. Then  $Y \in \wreach_r[v]$
 and so, in particular, both $\max X$ and $\max Y$ are in $ \wreach_r[v]$. It now follows from \cref{lem:ordering wreach} that $\max X\in \wreach_{2r}[\max Y]$.

Assume now that $(X,Y)$ is an info arc. Then $\max X \in Y$ or $\max Y \in X$. Without loss of generality assume the former. 
This implies that $\max X  \le \max Y$. Because $Y = \first_i(u)$ for some $u \in V(G)$ and some $i$, it holds that $Y \subset \wreach_r[u]$. By \cref{lem:ordering wreach}, it again follows that $\max X\in \wreach_{2r}[\max Y]$, as desired. 
\end{proof*}

\begin{claim}
\label{lem:path2_be}
Let $X$ be a non-root node of $B$ and $s\in \N$. Then for every non-root node $Y \in \wreach_s^{B,\le}[X]$, it holds that $\max Y \in \wreach_{2rs}^{G,\le}[\max X]$.
\end{claim}
\begin{proof*}
 Suppose $Y \in \wreach_s^{B,\le}[X]$ is a non-root node.
    Say this is witnessed by a path $X= Z_0,\dots,Z_{s'}=Y$ of length $s' \le s$ with $Y \le_B Z_i$ for all $i \in \{0,1,\dots,s'\}$; in particular, each $Z_i$ is non-root.
    Therefore, 
    $\max Y \le \max Z_i$ for all $i \in \{0,1,\dots,s'\}$.
    Since $Z_{i-1}$ and $Z_{i}$ are adjacent in~$B$,
    by \Cref{lem:path_be}, there is a path $\pi_i$ of length at most $2r$ from $\max Z_{i-1}$ to $\max Z_{i}$ in $G$
    with $\min(\max Z_{i-1}, \max Z_{i}) \le \min V(\pi_i)$.
    By concatenating the paths $\pi_1,\dots,\pi_{s'}$,
     we get a walk $\Pi$ of length at most $2rs' \le 2rs$ which starts in $\max X$ and ends in $\max Y$. 
    Since $Y=Z_{s'}$ is the $\le$-smallest node among $Z_0,\dots,Z_{s'}$, from the properties of paths $\pi_i$, it follows that $\max Y$ is the $\le$-smallest vertex on $\Pi$.
    The walk $\Pi$ therefore witnesses that $\max Y  \in \wreach_{2rs}^{G,\le}[\max X ]$, as desired.
\end{proof*}
Now \eqref{colbound_of_B_be} follows from \Cref{lem:rep_inverse_count_be} and \Cref{lem:path2_be}. The additional summand $1$ corresponds to taking into account also the root node of $B$. Since  \eqref{colbound_of_B_be} bounds the weak reachability number for any $s$ by a function of $s$ (here $r$ is fixed), \cref{lem:beBushes} follows.

% This finishes the proof of \cref{prop:imp}, without the `moreover' part.

%\begin{proof*}
%    Let $Y \in \wreach_q^{B,\le}[X]$.
%    This is witnessed by a path $X= Z_0,\dots,Z_{q'}=Y$ of length $q' \le q$ with $Y \le_B Z_i$ for all $i \in [q']$, and therefore also 
%    $\rep(Y) \le \rep(Z_i)$ for all $i \in [q']$.
%    Since $Z_{i-1}$ and $Z_{i}$ are adjacent in $B$,
%    by \Cref{lem:path} there is a path $\pi_i$ of length at most $2r$ from $\rep(Z_{i-1})$ to $\rep(Z_{i})$ in $G$ 
%    with $\min(\rep(Z_{i-1}), \rep(Z_{i})) \le \min V(\pi_i)$.
%    By concatenating the paths $\pi_1,\dots,\pi_{q'}$, we get a path $\Pi$ of length at most $2rq' \le 2rq$ in $G$ which starts in $\rep(X)$ and ends $\rep(Y)$. 
%    We claim that $\rep(Y)$ is the $\le$-smallest vertex on $\Pi$: if some vertex $w$ of $\Pi$ is such that $w < \rep(Y)$, then $w \in \pi_i$ for some $i$ and by \Cref{lem:path} 
%    $\rep(Z_{i-1}) \le w$ or $\rep(Z_i) \le v$, which means that $\rep(Z_{i-1}) < \rep(Y)$ or  $\rep(Z_i) < \rep(Y)$, a contradiction with $\rep(Y) \le \rep(Z_i)$ for all $i \in [q']$.
%    The path $\Pi$ therefore witnesses that $\rep(Y) \in \wreach_{2qr}^{G,\le}[\rep(X)]$ as desired.
%\end{proof*}
%Now (\ref{colbound_of_B}) follows from \Cref{lem:rep_inverse_count} and \Cref{lem:path2}.

\subsection{Reducing the number of labels}

We now proceed to prove \cref{lem:one label}.
Let $T$ be the tree of a bush $B$.
For every $a \in \Lambda$, let $T_a$ be the subtree of $T$ induced by all ancestors of leaves of $T$ labeled with $a$ (including those leaves).
For a node $v \in T$, we refer to the corresponding node in $T_a$ (if it exists) by $(v,a)$.
We will define a bush $B'$ with tree $T'$ as follows.
Let $T'$ be the tree constructed from the disjoint union of trees $\setof{T_a}{a\in \Lambda}$ by adding a fresh root and making it the parent of all the roots of trees $T_a$, $a \in \Lambda$.
Note that $\text{Leaves}(T) = \text{Leaves}(T')$. For the label set of $B'$ we take $\Lambda'=\{\bullet\}$, where $\bullet$ is a symbol.

%      Let $(v,a)$ and $(u,b)$ be two nodes in $T'$.
% We add an info arc $(e,a,b)$ between two nodes $(v,a)$ and $(u,b)$ in $T'$ if and only if there is an info arc $e$ between $v$ and $u$ in $T$.
% Let $x,y$ be two leaves in $T$ with lowest common info arc $e$ and $\lambda(x)=a$ and $\lambda(y)=b$.
% Note that the lowest common info arc between $(x,a)$ and $(y,b)$ in $T'$ is $(e,a,b)$.
% We set the label of this arc to be
% $$
% \lambda^{I'}((e,a,b)) = \begin{cases}
%     \{(\bullet,\bullet)\} & \quad \text{if }(a,b) \in \lambda^I(e), \\
%     \emptyset & \quad \text{otherwise}.
% \end{cases}
% $$
We add an info arc between two nodes $(u,a)$ and $(v,b)$ in $T'$ if and only if there is an info arc between $u$ and $v$ in $T$.
Let $x,y$ be two leaves in $T$ with lowest common info arc $(u,v)$ and $\lambda(x)=a$ and $\lambda(y)=b$.
Note that the lowest common info arc between $(x,a)$ and $(y,b)$ in $T'$ is $((u,a),(v,b))$.
We set the label of this arc to be
$$
\lambda^{I'}((u,a),(v,b)) = \begin{cases}
  \{(\bullet,\bullet)\} & \quad \text{if }(a,b) \in \lambda^I((u,v)), \\
  \emptyset & \quad \text{otherwise}.
\end{cases}
$$

      Thus, in the graph represented by $B'$ we consider two leaves adjacent if and only if their lowest common info arc is labeled with $\{(\bullet,\bullet)\}$. It is straightforward to verify that $B'$ represents the same graph as $B$.
  
      Let us now consider bushes as graphs consisting of both the tree- and the info-arcs.
      It is easy to see that $B'$ is a subgraph of the graph constructed by first taking the lexicographical product of $B$ and $K_{|\Lambda|}$
      and then adding a universal vertex adjacent to all other vertices (for the root of $T'$).
      Taking the lexicographical product increases all weak coloring numbers by a multiplicative factor at most $|\Lambda|$,
      and adding a universal vertex may additionally increase each of them by one.
  This proves \cref{lem:one label}.
    %\end{proof}

\subsection{Obtaining bushes by transductions}\label{sec:transduction section}
In this section, we discuss 
the following result, which is proved in the appendix.
\begin{theorem}\label{thm:trandsuction-thm}
  For every class $\DD$ that has structurally bounded expansion,
  there exist a class $\cal B$ of bushes that has bounded expansion and bounded depth, as well as a transduction $\trans B\from \DD\tto \cal B$ such that, for every $G\in\DD$, a bush $B_G$ representing $G$ 
  can be transduced by $\trans B$ from $G$; that is, $B_G\in\trans B(G)$.
\end{theorem}

As a corollary of \cref{thm:trandsuction-thm}, we may extend the result of \cref{thm:bush-be} to obtain analogous characterizations for classes that are \emph{structurally $\cal P$},
holds for other properties $\cal P$ of graph classes.
More precisely, let $\cal P$ be a property of graph classes satisfying the following conditions:
\begin{enumerate}[label=(P\arabic*)]
  \item\label{p1}
   every class $\DD$ with property $\cal P$ has bounded expansion,
  \item\label{p2} for every transduction $\trans T$ and class $\DD$ with property $\cal P$,
  if the class $\trans T(\DD)$ excludes some biclique as a subgraph, then the class $\trans T(\DD)$ has property $\cal P$.
\end{enumerate}
A simple example of such a property $\cal P$ is the property
of having bounded treedepth. Indeed, to see~\ref{p2} in this case, observe that transductions of bounded expansion classes have bounded shrubdepth, while classes of bounded shrubdepth that exclude some biclique as a subgraph in fact have bounded treedepth. An analogous reasoning shows that another example is the property of having bounded treewidth.
Yet another example is the property of having bounded \emph{sparse twin-width} \cite{BonnetGKTW21}, that is, having bounded twin-width and excluding some fixed biclique as a subgraph.
Note that the property of being nowhere dense satisfies condition \ref{p2}, but not condition \ref{p1}.

\begin{corollary}\label{cor:bushes}
  Let $\cal P$ be a property of graph classes satisfying conditions \ref{p1} and \ref{p2} above.
  The following conditions are equivalent for a class of graphs~$\DD$:
  \begin{itemize}[nosep]
   \item $\DD$ is obtained by a transduction of a class enjoying $\cal P$\\
   (we say that $\DD$ is \emph{structurally} $\cal P$),
   \item there are $d,\ell\in \N$ such that, for
   every $G\in \DD$, there is a bush $B_G$ representing $G$ where
   \begin{itemize}
       \item each bush $B_G$ 
       has depth at most $d$ and uses at most $\ell$ labels, and
       \item the class $\setof{B_G}{G\in \DD}$ has property $\cal P$.
   \end{itemize}   
   \end{itemize}
\end{corollary}
\begin{proof}
 It suffices to verify that the class $\BB=\setof{B_G}{G\in \DD}$ provided by \cref{thm:bush-be} enjoys property $\cal P$. We know that $\BB$ has bounded expansion, so it excludes some biclique as a subgraph. Further, by \cref{thm:trandsuction-thm}, $\BB$ can be transduced from a class enjoying $\cal P$. It follows from \ref{p2} that $\BB$ enjoys~$\cal P$. 
\end{proof}

\Cref{cor:bushes} in particular says that we can obtain 
combinatorial characterizations, completely analogous to~\cref{thm:bush-be},
of classes with structurally bounded treewidth
(which are exactly the stable classes of bounded cliquewidth~\cite{NesetrilMPRS21}),
or of classes with structurally bounded sparse twin-width
(which are exactly the stable classes of bounded twin-width~\cite{GajarskyPT21}).

%!TEX root = main.tex
\section{Quasi-bushes}\label{sec:quasi-bushes}

In this section, we provide a decomposition theorem for structurally nowhere dense classes of graphs. The decomposition notion --- {\em{quasi-bushes}} --- is similar in spirit to bushes, but there are important differences. While, in a bush, the info arcs connect pairs of nodes on the same level, in a quasi-bush, we use {\em{pointers}}: directed edges with tail in a leaf and head in an internal node of the quasi-bush. As before, the leaves of a quasi-bush are the vertices of the graph represented by it, and both leaves and pointers are labeled with a finite set of labels. The mechanics for encoding a graph in a quasi-bush is the following: to determine the adjacency between vertices $u$ and $v$, we find the lowest ancestor of $u$ that is pointed to by $v$ and inspect the pair of labels of $u$ and of the said~pointer. See \cref{fig:quasi-bush} for an illustration.
We proceed to formal details.

\begin{definition}
A quasi-bush consists of:
\begin{itemize}
\item a rooted tree $T$;
\item a set $D$ of directed edges (called {\em{pointers}}) from the leaves of $T$ to internal nodes of $T$ (we require that every leaf points to the root of $T$);
\item a labeling function $\lambda\from \leaves(T)\to \Lambda$, where $\Lambda$ is a finite set of \emph{labels}; %of the leaves of $T$ with elements of $\Lambda$;
\item a labeling function $\lambda^D\from D\to 2^{\Lambda}$.
\end{itemize}
A quasi-bush $B$ defines a directed graph $G(B)$ whose vertices are the leaves of $T$ and where the arc set is defined as follows: Let $u,v$ be two distinct
leaves and let $w$ be the lowest ancestor of $u$ such that $(v,w) \in D$. Then
$(u,v)$ is an arc in $G(B)$ if and only if $\lambda(u) \in \lambda^D((v,w))$.
\end{definition}

 As in the case of bushes, formally a quasi-bush $B$ represents a directed graph $G(B)$. If we want to represent an undirected graph $G$ by a bush $B$, we again require that $G(B)$ is the directed graph obtained from $G$ by replacing every edge with two oppositely-oriented~arcs. %FILL IN QUASI-BUSH PICTURE REFERENCE

% FILL IN QUASI-BUSH PICTURE
%\begin{figure}\centering
%  \includegraphics[page=1,scale=0.64]{pics}
%  \caption{A quasi-bush $B$ and the graph $G(B)$ defined by $B$.
  %TO UPDATE: In this bush, the leaves use one label $\bullet$ only,
  %which turns out to be always sufficient. 
  %There are two labels for the info arcs: red and blue (dotted). The color of an info loop at an inner node is indicated by the node's color. The vertices of $G(B)$ are the leaves of $B$, and two vertices $u,v$ are adjacent in $G(B)$ if and only if the lowest info arc (or info loop) above $u$ and $v$ is~red.
%  }
%  \label{fig:quasi-bush}
%\end{figure}

% 
% Note that, formally, a quasi-bush $B$ defines a directed graph, because the mechanism for recovering adjacency is not necessarily symmetric. By abuse of notation, we will usually denote by $G(B)$ also the undirected graph underlying $G(B)$, that is, where $u$ and $v$ are adjacent if and only if at least one of the arcs $(u,v)$ or $(v,u)$ is present in $G(B)$. In this way, a quasi-bush $B$ can also represent the undirected graph $G(B)$. Curiously, in our construction of quasi-bushes representing (undirected) graphs from structurally nowhere dense classes, recovering adjacency always works in a symmetric manner, making all arcs in the (directed) $G(B)$ bi-directional.

The Gaifman graph of a quasi-bush is the Gaifman graph of the structure consisting of $T$ and the edges from $D$. Whenever we speak about structural properties (nowhere denseness etc.) of a class $\BB$ of quasi-bushes, we do so with respect to the class of Gaifman graphs of quasi-bushes from $\BB$.

The main result of this section is the following:
\begin{theorem}
\label{thm:domainpreserving_snd_slightly_dense_trees}
Let $\CC$ be a nowhere dense class of graphs and let $\phi(x,y)$ be a formula. Then there exist $d,\ell\in \N$ and an almost nowhere dense class $\BB$ of quasi-bushes of depth at
most $d$ and with label set $\Lambda$ of size at most $\ell$ such
that, for every $G \in \trans I_\phi(\CC)$, there exists $B \in \BB$ with $G = G(B)$.\end{theorem}

Note that \Cref{thm:domainpreserving_snd_slightly_dense_trees} only speaks about interpretations that preserve the domain. We will generalize it to the more general setting (also of transductions) in \Cref{sec:closure}.

\Cref{thm:domainpreserving_snd_slightly_dense_trees} is a simple consequence of \Cref{thm:separator_trees} stated below. To phrase it, we will need the following definition.

\begin{definition}\label{def:quasi-bush}
Let $G$ be a graph. An {\em{$r$-separator quasi-bush}} for $G$ consists of:
\begin{itemize}
\item a rooted tree $T$ with $\leaves(T)=V(G)$;
\item a set $D$ of directed edges (called {\em{pointers}}) from the leaves of $T$ to internal nodes of $T$ (we require that every leaf points to the root of $T$); and
\item a function $\alpha$ which assigns to each internal node $w$ of $T$ a set $\alpha(w)$ such that the following holds: if $u$ and $v$ are two leaves of $T$ (vertices of $G$) and $w$ is the lowest ancestor of $v$ such that $(u,w) \in D$, then $\alpha(w)$ is an $r$-separator between $u$ and $v$ in $G$.
\end{itemize}
\end{definition}

\begin{theorem}
\label{thm:separator_trees}
Let $\CC$ be a nowhere dense class of graphs and $r \in \N$. Then there exist $d,m \in \N$ and an almost nowhere dense class $\BB$ of $r$-separator quasi-bushes of depth at most $d$ such that, for every $G \in \CC$, there exists $B \in \BB$ for which the function $\alpha$ satisfies $|\alpha(w)| \le m$ for each internal node $w$ of $T$.
\end{theorem}

With \cref{lem:separation and types}, \Cref{thm:domainpreserving_snd_slightly_dense_trees} follows quite easily from \cref{thm:separator_trees}.

\begin{proof}[Proof of \Cref{thm:domainpreserving_snd_slightly_dense_trees} using \Cref{thm:separator_trees}]
      Let $q$, $r$ be the numbers obtained by applying \Cref{lem:separation and types} to $\phi$.  Apply \Cref{thm:separator_trees} to 
      $\CC$ and $r$ to obtain $d$, $m$, and a class $\BB$ of $r$-separator quasi-bushes with the property claimed there.
      
      Consider $G\in \trans I_\phi(\CC)$, say $G=\trans I_\phi(H)$ for some $H \in \CC$. 
      By the assumed properties of $\BB$, there exists an $r$-separator quasi-bush $B^\circ \in \BB$ for $H$, say with the underlying tree $T$, pointer set $D$, and function $\alpha$, such that $B^\circ$ has depth at most~$d$ and $|\alpha(w)|\le m$ for every internal node $w$ of $T$. We have $\leaves(T)=V(G)$.
      We are going to construct a quasi-bush $B$ with $G=G(B)$. This quasi-bush will have $T$ as the underlying tree and pointer set $D$, so it remains to define the labeling functions $\lambda$ and $\lambda^D$.
      
      Let us fix an arbitrary order on the vertices of $G$. With a slight abuse of notation, for vertices $u,w \in V(G)$ we treat $u\alpha(w)$ as a tuple
      consisting of $\{u\} \cup \alpha(w)$ where $u$ comes first and then comes $\alpha(w)$ sorted in the specified order.
      For each leaf $v$, its label $\lambda(v)$ is defined to be the set $$\setof{(\tp^q_{H}(v\alpha(w)),i)}{\text{$w$ is an ancestor of $v$ at depth $i$ in $T$}}.$$
      The size of $\alpha(w)$ is bounded by $m$, which depends only on~$\CC$ and $r$, while $r$ depends only on $\phi$.
      Similarly, $q$ depends only on $\phi$ and the signature of $H$ depends only on $\CC$.
      This means that the number of $q$-types of ${\le}(m+1)$-tuples in $H$ is bounded by a constant depending only on~$\CC$ and $\phi$.
      Since the depth $d$ is also bounded, we conclude that $|\Lambda|$ depends only on $\CC$ and $\phi$, where $\Lambda$ is the codomain of $\lambda$.
      
      It remains to construct a function $\lambda^D \from D \to 2^\Lambda$ so that $G = G(B)$.
      Consider any $(u,w) \in D$ and let $i$ be the depth of $w$ in $T$.
      We define $\lambda^D((u,w))$ to be the set of all leaf-labels $A\in \Lambda$
      satisfying the following: there is $(\tau,i) \in A$ such that $\tau$ and $\tp^q_{H}(u\alpha(w))$ together imply $\phi(x,y)$ in the sense of \Cref{lem:separation and types}.
      
      We verify that indeed $G=G(B)$.
      Consider any $u,v \in V(G)$ and let $w$ be the lowest ancestor of $v$ such that $(u,w) \in D$.
      Then $\alpha(w)$ is an $r$-separator between $u$ and $v$ in $H$.
      By \Cref{lem:separation and types}, this means that $H\models \phi(u,v)$ (equivalently, $uv \in E(G)$) if and only if $\tp^q_{H}(v\alpha(w))$ and $\tp^q_{H}(u\alpha(w))$ together imply $\phi(x,y)$.
      On the other hand, $uv \in E(G(B))$ if and only if there is $(\tau,i) \in \lambda(v)$ such that $\tau$ and $\tp^q_{H}(u\alpha(w))$ together imply $\phi(x,y)$.
      At last, $(\tau,i) \in \lambda(v)$ if and only if $\tau = \tp^q_{H}(v\alpha(w))$.
      Therefore $uv \in E(G)$ if and only if $uv \in E(G(B))$, and we are done.
\end{proof}

\subsection{Strategy for the construction of quasi-bushes}

We slowly proceed to the proof of \Cref{thm:separator_trees}. We first discuss the intuition and, in particular, we explain why a direct lift of the reasoning from \Cref{sec:bushes} will not work.

While in bounded expansion graph classes, for every graph $G$ and every $r$, there is
an ordering $\le$ such that the weak reachability sets $\wreach_r[v]$ are
bounded in size by a function of $r$ only,
in nowhere dense graph classes, we can only claim a bound of the form $\Oh_\eps(n^\eps)$, where $n$ is the vertex count.
Therefore, if we repeated the arguments from \Cref{sec:bushes}, we would obtain bushes of depth $\Oh_\eps(n^\eps)$ and using $\Oh_\eps(n^\eps)$ labels. This is too much for our purposes: we would like to have decompositions with depth and label count bounded by constants depending only on the class.
In \Cref{sec:bushes}, we used prefixes of weak reachability sets as separators.
In this section, we explore ways to obtain similar separators in nowhere dense classes, but whose sizes are independent of~$n$.

The inspiration for our approach comes from the so-called \emph{Splitter game}~\cite{GroheKS2017}.
Given a graph $G$, radius $r$, and timeout~$\ell$,
the $(\ell,r)$-Splitter game is played between two players called \emph{Connector} and \emph{Splitter}, who take turns.
In the beginning, the arena is the whole graph $G$.
In the $i$th round, Connector chooses a vertex $v_i$ from the arena.
The arena is then restricted to the ball of radius $r$ with center at $v_i$.
Splitter then chooses a vertex $w_i$ which is removed from the arena.
Splitter wins the game if an empty arena is reached within $\ell$ rounds (see \cite[Definition 4.1]{GroheKS2017} for a precise definition.)
Grohe, Kreutzer and Siebertz showed that nowhere dense classes can be characterized using the Splitter game~\cite{GroheKS2017}.
\begin{theorem}[{\cite{GroheKS2017}}]\label{thm:splittergame}
  Let~$\CC$ be a nowhere dense class of graphs. Then, for
  every~$r>0$, there is~${\ell}>0$, such that for
  every~$G\in\CC$, Splitter has a strategy to win the~$({\ell},r)$-splitter game on~$G$.
\end{theorem}
Note that the number of rounds $\ell$ needed to win the game for Splitter depends only on $r$ and not on $n$.

Suppose that $\le$ is a total order on $V(G)$, and consider 
the strategy of Splitter where he always removes the smallest vertex from the current arena, and the strategy of Connector, where she always picks the same vertex $v$. Then it is not difficult to see that when the two players play according to this strategy, 
Splitter will remove the vertices of $\wreach^{G,\le}_r[v]$ one by one, from smallest to largest, until $v$ is finally removed.
Hence, if $G$ comes from a class with bounded expansion,
then we can find an order $\le$ such that 
this will last a bounded number of rounds only.
Moreover, the set of all possible plays (depending on the choice of $v$ by Connector) can be combined into a single tree, which will be essentially the tree of prefixes of weak $r$-reachability sets,
as constructed in \Cref{sec:bushes}.
Hence, the bush constructed in that section
is essentially a representation of the game tree corresponding to that particular strategy of Splitter. 

In the case of nowhere dense classes, there is no constant bound on the size of the weak $r$-reachability sets, but \cref{thm:splittergame}
says that there is an alternative strategy for Splitter that terminates in a bounded number of rounds.
The idea behind quasi-bushes is to construct a tree representing the possible plays according to that strategy.
In a certain sense,
the constant number of vertices $w_1,\dots,w_\ell$ played by Splitter to reach an empty arena (when Connector always plays the same vertex $v$) can act as a bounded-size substitute for $\wreach_r[v]$.
In this section, we follow Splitter's strategy to define subsets $M_r[v] \subseteq \wreach_r[v]$ with associated separators $S_r[v]$
whose sizes depend only on $r$.
To gain additional insights and streamline the construction, we will ``open the black box'' and base our reasoning on a particular strategy for Splitter proposed by Grohe, Kreutzer and Siebertz in their proof of \Cref{thm:splittergame}.

\subsection{Bounded-size separators}

We start by defining the sets $M_r[v]$ and $S_r[v]$.
%Roughly speaking, the set $M_r[v]$ will correspond to the choices of Connector and $S_r[v]$ to the choices of Splitter in the proof of \Cref{thm:splittergame}~\cite{GroheKS2017}.
In \Cref{lem:MrBasicProperties}, we verify their separator properties, while \Cref{lem:MrSmall} provides a bound on their sizes.

Fix a number $r\in\N$, a graph $G$, a total order $\le$ on the vertices of $G$, and a vertex $v\in V(G)$.
In the following sequence of steps, starting with step $k=1$,
we select vertices $m_1,m_2,\dots$ from $V(G)$, as well as paths $\pi_{ij}$ of length at most $2r$ between $m_i$ and $m_j$ in $G$, for all relevant $i<j$.

In the $k$th step, 
let $G_k$ be the graph obtained from $G$ by removing the vertices in $\bigcup_{1 \le i \le j < k} V(\pi_{ij})$.
In particular, $G_1=G$.
If $v \not\in V(G_k)$, then we terminate the construction.
Otherwise, let $m_k$ be the $\le$-smallest element of the ball of radius $r$ around $v$ in $G_k$.
For $1 \le i \le k$ let $\pi_{ik}$ be a path of length at most $2r$ in $G_i$ between $m_i$ and~$m_k$.
Such a path is guaranteed to exist in $G_i$, since both $m_i$ and $m_k$ are contained in the ball of radius $r$ around $v$ in $G_i$.
If there are multiple candidates for $\pi_{ik}$, it is crucial that the choice of $\pi_{ik}$ depends only on $G_i$, $m_i$, and $m_k$, and not on $v$.
(This can be done for example by choosing $\pi_{ik}$ minimal among the candidates with respect to the lexicographical ordering of its vertices in~$\le$.)

\begin{definition}
    Assume the process described above terminates after completing step $\ell$
    for a given number $r\in\N$, graph $G$ and order $\le$.
    For $v \in V(G)$ we define
    $$
    M^{G,\le}_r[v] = \{m_1,\dots,m_\ell\},
    $$
    $$
    S^{G,\le}_r[v] = \bigcup_{1 \le i \le j \le \ell} V(\pi_{ij}).
    $$
    Further, for each $0 \le k \le |G|$ define the {\em{$k$-prefixes}}  
    $$
    M^{k,G,\le}_r[v] = \{m_1,\dots,m_{\min(k,\ell)}\},
    $$
    $$
    S^{k,G,\le}_r[v] = \bigcup_{1 \le i \le j \le \min(k,\ell)} V(\pi_{ij}).
    $$
    When $G$ and $\le$ are clear from the context, we simply write $M_r[v]$, $M^k_r[v]$, $S_r[v]$ and~$S^k_r[v]$.
\end{definition}

The next lemma highlights 
three key properties of the sets defined above.
The crucial separator property follows from the third item:
if $k$ is minimal such that $S^{k}_{r}[v]$ $r$-separates $u$ and $v$, then
$S^{k-1}_{r}[v]$ does not $r$-separate $u$ and $v$, and hence $M^{k}_r[v] \subseteq \wreach_{2r}[u]$.
\begin{lemma}\label{lem:MrBasicProperties}
    Fix $r,k\in\N$ with $k\ge 1$, an order $\le$ on the vertices of a graph $G$, and vertices $u,v\in V(G)$. Then
    \begin{itemize}
        \item $M_r[v]\subset \wreach_r[v]$;
        \item $|S_r[v]| \le 2r|M_r[v]|^2$; and
        \item if $S^{k-1}_{r}[v]$ does not $r$-separate $u$ and $v$ in $G$, then $M^{k}_r[v] \subseteq \wreach_{2r}[u]$.
    \end{itemize}
\end{lemma}
\begin{proof}
    First, each vertex $m_k \in M_r[v]$ is the $\le$-smallest element of the ball of radius $r$ around $v$ in $G_k$.
    Hence $m_k \in \wreach_r^{G_k,\le}[v] \subseteq \wreach_r[v]$.

    Second,
    every path $\pi_{ij}$ has length at most $2r$ and endpoints $m_i,m_j \in M_r[v]$.
    So \[|S_r[v]| \le |M_r[v]| + {|M_r[v]| \choose 2} (2r-1)\le 2r|M_r[v]|^2.\]

    We proceed to the third item. Assume $S^{k-1}_{r}[v]$ does not $r$-separate $u$ and $v$ in $G$ (where $S^0_r[v]=\emptyset$).
    We can assume $M_{r}[v] = \{m_1,\dots,m_{\ell}\}$ with $k \le \ell$, since otherwise $v \in S^{k-1}_{r}[v]$ and $S^{k-1}_{r}[v]$ $r$-separates $v$ from every vertex.
    As $S^{k-1}_{r}[v]$ does not $r$-separate $u$ and $v$ in~$G$, 
    there exists a path $\rho$ of length at most $r$ from $u$ to $v$ in $G_{k}$.
    Then for every $1\leq k'\leq k$, $m_{k'}$ is the $\le$-smallest element of the ball of radius $r$ around $v$ in $G_{k'}$,
    and thus no vertex on $\rho$ is smaller in $\le$ than $m_{k'}$.
    This means there is a path of length at most $2r$ from $u$ to $v$ to $m_{k'}$ in $G$, and no vertex on this path is smaller than $m_{k'}$ in~$\le$. This
    witnesses that $m_{k'} \in \wreach_{2r}[u]$, implying that $M^k_r[v]\subseteq \wreach_{2r}[u]$.
\end{proof}

Next, \Cref{lem:MrSmall} below proves that in nowhere dense classes,
$|M_r[v]|$ (and in turn also $|S_r[v]|$) can be bounded by a function of $r$ only.
The argument closely follows the proof of \cite[Theorem 4.2]{GroheKS2017}
and uses the following characterization of nowhere dense classes through uniform quasi-wideness.
For a graph $G$ and $r\in \N$, 
a set $I \subseteq V(G)$ is a \emph{distance-$r$ independent set} if for all different $u,v\in I$, the distance between $u$ and $v$ in $G$ is larger than $r$. Note that a distance-$1$ independent set is just a standard independent~set.

\begin{definition}[Uniform Quasi-Wideness]\label{def:uqw}
    A class of graphs $\CC$ is \emph{uniformly quasi-wide} if for every $r \in \N$ there exists
    a function $N_r \from \N \to \N$ and a constant $s_r \in \N$ such that for all $m \in \N$, $G \in \CC$, and $A \subseteq V(G)$
    with $|A| \ge N_r(m)$, there exists $S \subseteq V(G)$ with $|S| < s_r$ and $I \subseteq A \setminus S$ with $|I| \ge m$ such that $I$
    is a distance-$r$ independent set in~$G-S$.
\end{definition}
\begin{theorem}[{\cite{nevsetvril2011nowhere}}]\label{lem:uqw}
    A class $\CC$ of graphs is nowhere dense if and only if it is uniformly quasi-wide.
\end{theorem}

\begin{lemma}\label{lem:MrSmall}
  Fix a nowhere dense class $\CC$.
  Then for every $r\in\N$, there is a number $\ell$ 
  such that, for every $G\in\CC$, order $\le$ on $V(G)$, and $v\in V(G)$, we have $|M_r[v]|\le \ell$.
\end{lemma}
\begin{proof}
    Let $r \in \N$.
    By \Cref{lem:uqw}, since $\CC$ is nowhere dense, it is also uniformly quasi-wide. Let $N=N_r \from \N \to \N$ and $s=s_r \in \N$ be the function and constant witnessing uniform quasi-wideness for radius $r$.
    We set $\ell = N(2s)-1$.
    Assume for contradiction that
    there exist $G\in\CC$ with order $\le$ on $V(G)$ and $v\in V(G)$
    such that $A = M_r[v] = \{m_1,\dots,m_k\}$ for some $k > \ell$.
    Since $|A|\ge N(2s)$, by uniform quasi-wideness there exists a set $S \subseteq V(G)$ with $|S| < s$ 
    and $I \subseteq A \setminus S$ with $|I| = 2s$ such that $I$ is a distance-$r$ independent set in $G \setminus S$.

    Write $I$ as $I=\{m_{a_1},\dots,m_{a_{2s}}\}$ with $a_1 < \dots < a_{2s}$.
    For all $1 \le i \le s$,
    $V(\pi_{a_{2i-1},a_{2i}}) \cap S \neq \emptyset$,
    since there is no path of length at most $r$ between $m_{a_{2i-1}}$ and $m_{a_{2i}}$ in $G \setminus S$.
    Also, for all $1 \le i<j \leq s$ it holds
    $V(\pi_{a_{2i-1},a_{2i}}) \subseteq S_{a_{2i}}[v]$ and $V(\pi_{a_{2j-1},a_{2j}}) \cap S_{a_{2i}}[v] = \emptyset$.
    Therefore $V(\pi_{a_{2i-1},a_{2i}}) \cap V(\pi_{a_{2j-1},a_{2j}})  = \emptyset$.
    This implies $|S| \ge s$, a contradiction to the assumption $|S| < s$.
\end{proof}

\subsection{Construction of separator quasi-bushes}

We now show how to construct an $r$-separator quasi-bush for a graph equipped with a total order on its vertices. 
\begin{definition}
    Let $r\in \N$ and $G$ be a graph with an order $\le$ on its vertices.
    We will treat the sets $M^{k}_r[v]$ as tuples $(m_1,m_2,\dots)$ by ordering their elements in the order of discovery.
    We define the $r$-separator quasi-bush $\BS(G,\le)$ as follows.
    \begin{itemize}
        \item 
            The underlying tree $T$ consists of the leaves $V(G)$ and the internal nodes $\setof{M_r^k[v]}{v \in V(G), k \in \N}$.
            The root of $T$ is the empty tuple.
            The parent of an internal node $(m_1,\dots,m_\ell)$ is the node $(m_1,\dots,m_{\ell-1})$.
            The parent of a leaf $v$ is $M_r[v]$.
        \item $D$ comprises pointers $(u,\emptyset)$ for all $u\in V(G)$ (note that $\emptyset$ is the root of $T$) and pointers $(u,M^k_r[v])$ for all $u,v\in V(G)$ and $k\ge 1$ such that  $S^{k-1}_{r}[v]$ does not $r$-separate $u$ and $v$ in $G$.
        \item For each internal node $M_r^k[v]$ of $T$, we set $\alpha(w) = S_r^k[v]$.
    \end{itemize}
\end{definition}
We will use capital letters $X,Y,Z,\ldots$ to denote the nodes of $\BS(G,\le)$.
We start by showing that $\BS(G,\le)$ is indeed an $r$-separator quasi-bush for $G$.

\begin{lemma}\label{lem:doesSeparate}
    Let $G$ be a graph with order $\le$ on its vertices and $r \in \N$.
    Then $\BS(G,\le)$ is an $r$-separator bush for $G$.
\end{lemma}
\begin{proof}
  The only non-trivial property to check is the last item of \Cref{def:quasi-bush}.
  Let $u,v \in V(G)$ and $X$ be the lowest ancestor of $v$ in $\BS(G,\le)$ such that $(u,X) \in D$. 
  We show that $\alpha(X)$ is an $r$-separator between $u$ and $v$ in $G$.
  %It is then obvious that $\BS(G,\le)$ also satisfies the remaining properties for being an $r$-separator bush for $G$.

  If $X$ is the parent of $v$
  then $X=M_r[v]$ and $\alpha(X)=S_r[v]$. Since $v \in S_r[v]$, 
  $\alpha(X)$ is an $r$-separator between $u$ and $v$.
  Thus, we can assume from now on that $X$ is not the parent of~$v$.
  Let $Y$ be the ancestor of $v$ that is a child of $X$.
  We have $X=M_r^k[v]$ and $Y=M_r^{k+1}[v]$ for some $k \in \N$.
  Since $X$ is the lowest ancestor of $v$ with $(u,X)\in D$, we have that $(u,M_r^{k+1}[v]) \not\in D$.
  Then by the definition of $D$ and contrapositive, $\alpha(X) = S^{k}_{r}[v]$ $r$-separates $u$ and $v$.
\end{proof}

It turns out that if $G$ is sparse, so is $\BS(G,\le)$.
The proof of the next lemma follows the same lines as the proof of \cref{lem:beBushes}.
\begin{lemma}\label{lem:slightlyDenseBushes}
    Let $\CC$ be a nowhere dense graph class and $r \in \N$.
    Then, for every $G\in \CC$, there is an order $\le$ on the vertices of $G$ such that the class $\{\BS(G,\le)\colon G\in \CC\}$ is almost nowhere~dense.
\end{lemma}
\begin{proof}
    Let $G \in \CC$.
    By \Cref{cor:universalND}, there is a universal order $\le$ such that
    ${\wcol_{q}(G,\le)} \le \Oof_{\CC,\varepsilon,q}(|G|^\varepsilon)$ for all $\varepsilon > 0$, $q \in \N$.
    We verify that fixing this order for each $G\in \CC$ satisfies the premise of the lemma.
    
    Let $B = \BS(G, \le)$ be the $r$-separator quasi-bush for $G$ constructed for the order~$\le$. 
    Suppose $B$ has underlying tree $T$, pointer set $D$, and separators $\alpha$. For a nonempty set $X\subseteq V(G)$, by $\min X$ and $\max X$ we denote the $\le$-minimum and $\le$-maximum element of $X$, respectively. Further, with each non-root node $X$ of $T$ with associate its {\em{representative}} $\rep(X)\in V(G)$ as follows:
    \begin{itemize}[nosep]
     \item If $X$ is an internal node, then $\rep(X)=\max X$.
     \item If $X$ is a leaf, then $\rep(X)=X$.
    \end{itemize}
    We extend the order $\le$ from $V(G)$ to all nodes of $B$ so that the root of $B$ comes first and for any two non-root nodes $X,Y$ of $B$ with $\rep(X) \le \rep(Y)$ it holds that $X \le Y$.
    
    In the remainder of the proof we shall show that there exists a constant $c$, depending only on $\CC$ and $r$
    such that for all $q\in \N$ we have
    \begin{equation}\label{colbound_of_B}
    \wcol_q(B,\le) \le 1+\wcol_{2qr}(G,\le) \cdot (\wcol_{2r}(G,\le)^c+1).
    \end{equation}
    That $\{\BS(G, \le)\colon G\in \CC\}$ is almost nowhere dense then follows from \Cref{cor:universalAlmostND} and the assumption that $\CC$ is nowhere dense.
    
    We will need the following observations.
    \begin{claim}
    \label{lem:rep_inverse_count}
    There exists a constant $c$ depending only on $\CC$ and $r$ such that
    $|\setof{Y \in V(T)}{\rep(Y)=v}| \le |\wreach^{G,\le}_{2r}[v]|^c+1$ for all $v\in V(G)$.
    \end{claim}
    \begin{proof*}
    Clearly there is only one leaf $Y$ of $T$ such that $\rep(Y) = v$, namely $Y=v$.
    In the rest of the proof we bound the number of internal nodes $Y$ with $\rep(Y) = v$.
    %that $M_r^k[u] \subset \wreach^{G,\le}_r[u]$
    For such a node $Y$ it holds that $Y = M_r^k[u]$ for some $u \in V(G)$ and $k \in \N$.
    By \Cref{lem:MrBasicProperties}, we have $Y \subseteq \wreach^{G,\le}_r[u]$.
    As $v=\max Y$, it follows that each $w \in M_r^k[u]$ is weakly reachable from $v$ by a path of length at most $2r$ through $u$.
    In other words, $Y \subseteq \wreach^{G,\le}_r[u] \subseteq \wreach^{G,\le}_{2r}[v]$.
    By \Cref{lem:MrSmall}, we can choose $c$, depending only on $\CC$ and $r$, so that $|Y| \le c$.
    Then the number of internal nodes $Y$ with $v=\rep(Y)$ is bounded by $|\wreach_{2r}[v]|^c$.
    \end{proof*}
    
    \begin{claim}
    \label{lem:path}
    Let $X,Y$ be two non-root nodes of $B$ that are adjacent, either in $T$ or via a pointer. Then in $G$ there exists a path $\pi$ of length at most $2r$ with endpoints $\rep(X)$ and $\rep(Y)$ such that 
    $$\min(\rep(X), \rep(Y)) \le \min V(\pi).$$
    \end{claim}
    \begin{proof*}
    We prove that, in each case, it holds that $\rep(X)\in \wreach_{2r}[\rep(Y)]$ or $\rep(Y)\in \wreach_{2r}[\rep(X)]$. Then any path $\pi$ witnessing this weak reachability satisfies the premise of the claim.
    
    Assume first $Y$ is the parent of $X$. 
    We consider two subcases: either $X$ is an internal node or a leaf of $T$.
    
    If $X$ is a leaf of $T$, then $X=\rep(X) = v$ for some $v \in V(G)$, and $Y = M_r[v]$.
    By \Cref{lem:MrBasicProperties}, $\rep(Y)=\max Y$ is weakly $r$-reachable from $\rep(X)=v$.
    
    If $X$ is an internal node of $T$, then we have $Y = M_r^{k-1}[u] =  \set{m_1,\ldots, m_{k-1}}$ and $X = M_r^{k}[v] =\set{m_1,\ldots, m_k}$ for some $u,v\in V(G)$ and $k\ge 2$. In this case  $\rep(X) = m_k$  and $\rep(Y) = m_{k-1}$. By Lemma~\ref{lem:MrBasicProperties}, $M_r^k[v]\subset \wreach_r[v]$. This means $\rep(Y)=m_{k-1}$ is weakly $2r$-reachable from $\rep(X)=m_k$  through $v$.
    
    Finally, we are left with the case when $(X,Y)$ is a pointer. Then $X = u$ and $Y = M_r^k[v]$ for some $u,v \in V(G)$ and $k\ge 1$, and $S_r^{k-1}[v]$ does not $r$-separate $u$ and $v$ in~$G$. 
    We have $\rep(X)=u$ and
    by Lemma~\ref{lem:MrBasicProperties}, it holds that $\rep(Y) \in M^{k}_r[v] \subseteq \wreach_{2r}[u]$.
    So $\rep(Y) \in \wreach_{2r}[\rep(X)]$ as desired.
    \end{proof*}
    
    \begin{claim}
    \label{lem:path2}
    Let $X$ be a node of $B$ and $q\in \N$. Then for every non-root node $Y \in \wreach_q^{B,\le}[X]$ it holds that $\rep(Y) \in \wreach_{2rq}^{G,\le}[\rep(X)]$.
    \end{claim}
    \begin{proof*}
        Suppose $Y \in \wreach_q^{B,\le}[X]$ is a non-root node.
        Say this is witnessed by a path $X= Z_0,\dots,Z_{q'}=Y$ of length $q' \le q$ with $Y \le_B Z_i$ for all $i \in \{0,1,\dots,q'\}$; in particular, each $Z_i$ is non-root.
        Therefore,  
        $\rep(Y) \le \rep(Z_i)$ for all $i \in \{0,1,\dots,q'\}$.
        Since $Z_{i-1}$ and $Z_{i}$ are adjacent in~$B$,
        by \Cref{lem:path} in $G$ there is a path $\pi_i$ of length at most $2r$ from $\rep(Z_{i-1})$ to $\rep(Z_{i})$
        with $\min(\rep(Z_{i-1}), \rep(Z_{i})) \le \min V(\pi_i)$.
        By concatenating the paths $\pi_1,\dots,\pi_{q'}$, we get a walk $\Pi$ of length at most $2rq' \le 2rq$ which starts in $\rep(X)$ and ends in $\rep(Y)$. 
        Since $Y=Z_{q'}$ is the $\le$-smallest node among $Z_0,\dots,Z_{q'}$, from the properties of paths $\pi_i$ it follows that $\rep(Y)$ is the $\le$-smallest vertex on $\Pi$.
        The walk $\Pi$ therefore witnesses that $\rep(Y) \in \wreach_{2qr}^{G,\le}[\rep(X)]$, as desired.
    \end{proof*}
    Now (\ref{colbound_of_B}) follows from \Cref{lem:rep_inverse_count} and \Cref{lem:path2}. Note that the additional summand $1$ corresponds to taking into account also the root node of $T$.
    \end{proof}

At last, we can prove \Cref{thm:separator_trees}.

\begin{proof}[Proof of Theorem~\ref{thm:separator_trees}]
    By \Cref{lem:slightlyDenseBushes},
    there exists an almost nowhere dense class $\BB$ such that,
    for every $G \in \CC$, there exists an ordering $\le$ 
    with $\BS(G,\le) \in \BB$.
    By \Cref{lem:doesSeparate}, $\BS(G,\le)$ is an $r$-separator quasi-bush of $G$.
    The depth of $\BS(G,\le)$ is bounded by \Cref{lem:MrSmall} and the size of the sets $\alpha(w)$ is bounded by the second item of \Cref{lem:MrBasicProperties}.
\end{proof}

%!TEX root = main.tex

\section{Closure in monadically NIP classes}\label{sec:closure}

In this section, we lift results from domain-preserving interpretations to transductions.
The crucial difference is that transductions may remove vertices.
Specifically, we show that, from \Cref{thm:domainpreserving_snd_slightly_dense_trees}, we can derive \Cref{thm:bush-nd}.

We first discuss the key difficulty in \Cref{thm:bush-nd} in comparison to the setting of \Cref{thm:domainpreserving_snd_slightly_dense_trees}.
Assume we are given a huge graph $G$, a formula  $\psi(x,y)$ used for the interpretation, and a small set $A \subseteq V(G)$ to which we will restrict the interpreted graph at the end. \Cref{thm:domainpreserving_snd_slightly_dense_trees} provides a quasi-bush $B$ representing $\trans I_\psi(G)$ such that $B$ belongs to a fixed almost nowhere dense class $\BB$. This means that the weak coloring numbers of $B$ are bounded by $\Oh_{\eps}(|G|^\eps)$. If we would like to obtain a quasi-bush representing the induced subgraph $\trans I_\psi(G)[A]$, we can restrict $B$ to the vertices of $A$ and their ancestors. However, for the weak coloring numbers of the restricted bush, we can only claim the inherited upper bounds of the form $\Oh_\eps(|G|^\eps)$, while \Cref{thm:bush-nd} would postulate  bounds of the form $\Oh_\eps(|A|^{\eps})$, that is, dependent only on the vertex count of the final graph output by the~transduction.

To fix this problem, we prove an auxiliary statement which roughly says the following: if a graph $H$ can be transduced from a graph $G$, then $H$ can be also transduced from an induced subgraph $\hat G$ of $G$ whose size is bounded in terms of the size of $H$. The latter transduction can be different (slightly more complicated) than the original one. The dependence of the size of $\hat G$ on the size of $H$ can be non-elementary in general, but we show that it is polynomial provided $G$ belongs to a fixed class that is monadically NIP (\Cref{thm:NIPLoewenheimSkolem}), and even of the form $\Oh(|H|^{1+\eps})$ for any $\eps>0$ provided $G$ belongs to any fixed nowhere dense class (\Cref{thm:NWDLoewenheimSkolem}). This means that in the scheme proposed in the previous paragraph, we may assume that $|G|$ is bounded polynomially in $|A|$, so the bounds $\Oh_\eps(|G|^\eps)$ and $\Oh_\eps(|A|^{\eps})$ are equivalent.

% will add some colors to $G$ and slightly change the formula of the interpretation.
% The upcoming \Cref{lem:subgraph} shows that it is then enough to consider a subgraph $G'$ of $G$, whose size is bounded in $A$.
% In general, this bound may be non-elementary.
% In the subsequent \Cref{thm:NIPLoewenheimSkolem}, we bound $G'$ by $|A|^{O(1)}$ for monadically NIP classes,
% by using their ???-Property.
% While this is enough for our use case,
% it may be of independent interest that for nowhere dense classes, one may bound
% $G'$ by $O(|A|^{1+\eps})$ for every $\eps > 0$
% (\Cref{lem:NWDLoewenheimSkolem}).

We proceed to formal details.
First, we need some definitions.
Consider a first-order formula $\phi(\tup x, \tup y)$ with a partitioning of its free variables into $\tup x$ and $\tup y$,
a graph $G$, a tuple $\tup u \in V(G)^{\tup y}$ and a set $A \subseteq V(G)$.
The \emph{$\phi$-type} of $\tup u$ over $A$, denoted by $\tp^\phi_G(\tup u/A)$,~is
\[
\tp^\phi_G(\tup u/A) = \{ \tup v \in A^{\tup y} \colon G \models \phi(\tup u,\tup v)\}.
\]
% Moreover, for a number $q \in \N$, a graph $G$, a tuple $\tup u\in V(G)^{\tup x}$ where $\tup x$ is a tuple of variables, and a set $A \subseteq V(G)$, we define
% \[
% \tp^q_G(\tup u/A) = \left( \tp^\phi_G(\tup u/A)\colon \phi(\tup x,\tup y)\right),
% \]
% where $\phi(\tup x,\tup y)$ ranges over all formulas of quantifier rank $q$ with a partition of the free variables where the first part is~$\tup x$.
We omit the subscript if the graph is clear from the context.

Our main tool in this section applies in the more general setting of monadically NIP classes.
A graph class $\CC$ is \emph{monadically NIP} if, for every transduction $\trans T$, $\trans T(\CC)$ is not the class of all graphs.
For example, structurally nowhere dense classes are monadically NIP.
Suppose a graph class $\CC$ is monadically NIP. Then, 
as a consequence of the Sauer-Shelah Lemma \cite{sauer1972density, shelah1972combinatorial},
for every formula $\phi(\tup x,y)$, there exist
constants $c, d \in \N$ such that,for every $G \in \CC$ and $A \subseteq V(G)$, we have 
$$
\left|\{ \tp^{\psi}_G(u/A) \colon u \in V(G) \}\right | \le c \cdot |A|^{d}.
$$
Using this, we prove the following.

\begin{theorem}\label{thm:NIPLoewenheimSkolem}
    Let $\CC$ be a monadically NIP graph class and let $\phi(\tup x)$ be a formula.
    Then there exist a formula $\hat\phi(\tup x)$ and $c, k \in \N$
    such that, for all $G \in \CC$, $A \subseteq V(G)$ there exists 
    a monadic lift $\hat G$ of an induced subgraph of $G$ such that \[|\hat G| \le c |A|^{k}\quad\textrm{and}\quad
    \phi(G)[A] = \hat \phi(\hat G)[A].\]
\end{theorem}

For nowhere dense classes, Pilipczuk et al.~\cite{numberOfTypes} gave much stronger bounds on the number of types.
%Firstly, let us recall the stronger bound on the number of types in nowhere dense classes, proved in~\cite{numberOfTypes}.
\begin{theorem}[\cite{numberOfTypes}]\label{thm:fewTypes}
Let $\CC$ be a nowhere dense graph class and $\phi(\tup x, \tup y)$ be a first-order formula.
For every $\eps > 0$, there exists a constant $c$ such that for every
$G \in \CC$ and every nonempty $A \subseteq V(G)$, it holds~that
$$
|\{ \tp^\phi_G(\tup u/A) \colon \tup u \in V(G)^{\tup x} \}| \le c \cdot |A|^{|\tup x|+\eps}.
$$
\end{theorem}
Using these, we can prove the following strengthening of \cref{thm:NIPLoewenheimSkolem} for nowhere dense classes.

\begin{theorem}\label{thm:NWDLoewenheimSkolem}
    Let $\CC$ be a nowhere dense graph class, $\phi(\tup x)$ be a formula, and $\eps > 0$.
    Then there is a formula $\hat \phi(\tup x)$ (depending only on $\phi(\tup x)$) and a constant $c$
    such that for all $G \in \CC$, $A \subseteq V(G)$, there exists 
    a monadic lift $\hat G$ of an induced subgraph of $G$ such~that \[|\hat G| \le c |A|^{1+\eps}\quad \textrm{and}\quad \phi(G)[A] = \hat \phi(\hat G)[A].\]
\end{theorem}

\cref{thm:NIPLoewenheimSkolem,thm:NWDLoewenheimSkolem} are immediate consequences of the following.
\begin{lemma}\label{lem:subgraph}
  Let $\CC$ be a graph class and $\phi(\tup x)$ be a formula of length
  $\ell$ and quantifier rank $q$.
  Assume there exist $c, \eps > 0$ such that
  for all formulas $\psi(\tup x,y)$ of length at most $\ell$
  and quantifier-rank at most $q$,
  and all $G \in \CC$ and $A \subseteq V(G)$,
  $$
  \left|\left\{ \tp^{\psi}_G(u/A) \colon u \in V(G) \right\}\right| \le c \cdot |A|^{1 + \eps}.
  $$
  Then there exists a formula $\hat\phi(\tup x)$ (depending only on $\phi(\tup x)$)
  such that for all $G \in \CC$, $A \subseteq V(G)$ there exists 
  a monadic lift $\hat G$ of an induced subgraph of $G$ such that 
  \[|\hat G| \le \ell^q (1+c)^{\frac{(1+\eps)^q-1}{\eps}}|A|^{(1+\eps)^{q}}\quad \textrm{and}\quad  \phi(G)[A] = \hat\phi(\hat G)[A].\] 
\end{lemma}
\begin{proof}
  We prove the lemma by induction on $q$.
  If $q=0$ the lemma holds,
  because for quantifier-free formulas $\phi(\tup x)$ we have $\phi(G)[A] = \phi(G[A])[A]$, so we may take $\hat G=G[A]$.

  Assume then that $q>0$.
  There is a set $\Psi$ of formulas with quantifier rank at most $q-1$
  such that $\phi(\tup x)$ is a Boolean combination of formulas $\exists y\,
  \psi(\tup x,y)$, $\psi \in \Psi$. Note that we may choose $\Psi$ so that $|\Psi|\leq \ell$ and each formula in $\Psi$ has length at most $\ell$.

  Let us fix a graph $G \in \CC$, a set $A \subseteq V(G)$, and a formula $\psi \in \Psi$. 
  By assumption, we may find a set $A_\psi \subseteq V(G)$ with $|A_\psi| \le c \cdot |A|^{1+\eps}$ such that
  $$
  \{ \tp^{\psi}_G(u/A) \colon u \in V(G) \} =
  \{ \tp^\psi_G(u/A) \colon u \in A_\psi \}.
  $$
  Let us further fix a tuple $\tup v \in A^{\tup x}$.
  If $G \models \exists y\, \psi(\tup v,y)$,
  then there exists a witness $u' \in V(G)$ such that $G \models \psi(\tup v,u')$.
  Note that we can replace $u'$ with $u \in A_\psi$ such that $\tp^{\psi}_G(u/A) = \tp^{\psi}_G(u'/A)$, and we still have $G \models \psi(\tup v,u)$.
  This means that
  \begin{equation}\label{eq:step1}
  G \models \exists y\, \psi(\tup v,y)\ \Leftrightarrow\
  G \models \psi(\tup v,u) \text{ for some } u \in A_\psi.
  \end{equation}
  The formula $\psi(\tup x,y)$ has quantifier rank at most $q-1$ and length at most $\ell$.
  By the induction assumption,
  there exists a formula $\hat\psi(\tup x)$
  a monadic lift $\hat G$ of an induced subgraph of $G$ such that 
  $|\hat G_\psi| \le \ell^{q-1} (1+c)^{\frac{(1+\eps)^{q-1}-1}{\eps}}(|A|+|A_\psi|)^{(1+\eps)^{q-1}}$
  and
  \[\psi(G)[A \cup A_\psi] = \hat\psi(\hat G_\psi)[A \cup A_\psi].\]
  Therefore
  \begin{equation}\label{eq:step2}
      \text{ for all } u \in A_\psi,
      G \models \psi(\tup v,u) \Leftrightarrow
      \hat G_\psi \models \hat\psi(\tup v,u).
  \end{equation}
  Combining (\ref{eq:step1}) and (\ref{eq:step2}) yields
  \begin{equation}\label{eq:step3}
  G \models \exists y\, \psi(\tup v,y) \Leftrightarrow
  \hat G_\psi \models \hat\psi(\tup v,u) \text{ for some } u \in A_\psi.
  \end{equation}
  We may assume that all graphs $\hat G_\psi$, $\psi \in \Psi$, add distinct colors to $G$.
  Let $\hat G$ be the monadic lift of $G[\bigcup_{\psi \in \Psi}V(\hat G_\psi)]$
  that contains all the colors of $\hat G_\psi$ as well as two new relations
  $P_\psi^{\hat G} = A_\psi$ and $Q_\psi^{\hat G} = V(\hat G_\psi)$, for every $\psi \in \Psi$.
  We construct $\psi'(\tup x,y)$ from $\hat\psi(\tup x,y)$ by relativizing all quantifiers to $Q_\psi$, i.e., 
  replacing all quantifiers $\exists z\, \xi$ in $\psi$ with $\exists z\, Q_\psi(z) \land \xi$.
  Thus
  \begin{equation}\label{eq:step4}
  \hat G_\psi \models \hat\psi(\tup v,u) \text{ for some } u \in A_\psi \Leftrightarrow
  \hat G \models \exists y\, P_\psi(y) \land \psi'(\tup v,y).
  \end{equation}
  Remember that $\phi(\tup x)$ is a Boolean combination of formulas $\exists y\, \psi(\tup x,y)$, $\psi \in \Psi$.
  At last, we construct $\hat\phi(\tup x)$ from $\phi(\tup x)$ by replacing each such subformulas
  $\exists y\, \psi(\tup x,y)$ with $\exists y P_\psi(y) \land \psi'(\tup x,y)$.
  Then (\ref{eq:step3}) and (\ref{eq:step4}) together imply that 
  $G \models \phi(\tup v) \Leftrightarrow \hat G \models \hat\phi(\tup v)$.
  Since this holds for every $\tup v \in A^{\tup x}$, we have
  $\phi(G)[A] = \hat\phi(\hat G)[A]$.
  Finally, recalling that $|\Psi|\le \ell$,
  \begin{align*}
      |\hat G| & \le \sum_{\psi\in\Psi} |\hat G_\psi| \\
           & \le \sum_{\psi\in\Psi} \ell^{q-1} (1+c)^{\frac{(1+\eps)^{q-1}-1}{\eps}}(|A|+|A_\psi|)^{(1+\eps)^{q-1}}\\
           & \le \ell^{q} (1+c)^{\frac{(1+\eps)^{q-1}-1}{\eps}} ((1+c) |A|^{1+\eps})^{(1+\eps)^{q-1}} \\
           & =   \ell^{q} (1+c)^{\frac{(1+\eps)^{q}-1}{\eps}} |A|^{(1+\eps)^{q}}. \qedhere
  \end{align*}
\end{proof}

\Cref{thm:bush-nd} now follows from \Cref{thm:domainpreserving_snd_slightly_dense_trees} and either of \cref{thm:NIPLoewenheimSkolem} or \cref{thm:NWDLoewenheimSkolem}.
%see \cref{app:closure} for a proof.

\begin{proof}[Proof of \Cref{thm:bush-nd} using \Cref{thm:domainpreserving_snd_slightly_dense_trees}]
  By the assumption, there exist a nowhere dense graph class $\CC$ and a transduction $\trans T$ such that $\DD \subseteq \trans T(\CC)$. 
  By \Cref{lem:ND-nocopy}, we can assume $\trans T$ to be non-copying.
  
  Further, the subgraph closure of a nowhere dense class is again nowhere dense. So we may assume that $\CC$ is closed under taking subgraphs.
  
  Fix any graph $H\in \DD$. 
  Then there is a graph $G\in \CC$ and $A\subseteq V(G)$ such that $H=\trans I_\psi(G)[A]$, where $\psi(x,y)$ is the formula used in $\trans T$.
  Let $\hat\psi(x,y)$ and $c, k \in \N$ be the formula and the constants obtained by applying \Cref{thm:NIPLoewenheimSkolem} to $\CC$ and~$\psi$.
  By \Cref{thm:NIPLoewenheimSkolem}, there exists
  a monadic lift $\hat G$ of an induced subgraph of $G$ such that $|\hat G| \le c |H|^{d}$
  and $H =\trans I_{\hat\psi}(\hat G)[A]$. Note that as $\CC$ is closed under taking subgraphs, we have~$\hat G\in \CC$.
  
  Apply \Cref{thm:domainpreserving_snd_slightly_dense_trees} to $\CC$ and $\hat \psi$ to obtain 
  $d, \ell$ and an almost nowhere dense class $\BB^\circ$ of quasi-bushes, each of depth at
  most $d$ and using a label set $\Lambda$ of size at most $\ell$.
  As $\hat G\in \CC$, there exists $B \in \BB^\circ$ such that $\trans I_{\hat\psi}(\hat G) = G(B)$.
  This means we have a quasi-bush $B$ for $\trans I_{\hat\psi}(\hat G)$, but want a quasi-bush $B'$ for the subgraph $\trans I_{\hat\psi}(\hat G)[A]$.
  Such a quasi-bush $B'$ is readily obtained by removing from $B$ all leaves that are not contained in $A$ and all internal nodes without any descendant in $A$.

  We have $|B| \le d|\hat G| \le cd|A|^k \le cd|B'|^k$. Since, $\CC$ is almost nowhere dense, for every $r\in \N$ and $\eps>0$ we have $\wcol_r(B)\leq \Oh_{r,\eps}(|B|^\eps)$. Therefore,
  \[\wcol_r(B')\leq \wcol_r(B)\leq \Oh_{r,\eps}(|B|^\eps)\leq \Oh_{r,\eps}(|B'|^{k\eps}).\]
  By rescaling $\eps$ by factor $k$ we conclude that $\wcol_r(B')\leq \Oh_{r,\eps}(|B'|^{\eps})$. Hence, the class $\BB$ of all quasi-bushes $B'$ obtained as above is almost nowhere dense, as intended.

  The proof that one label for the leaves and two labels for the pointers suffices is completely analogous to the proof in the case of bushes (see \cref{lem:one label}).
\end{proof}

%!TEX root = main.tex
\section{Low shrubdepth covers}
\label{sec:lsd}
% \begin{manualtheorem}{\ref{thm:lsd-nd}}
%  Let $\DD$ be a structurally nowhere dense class of graphs and $p\in \N$ and $\eps>0$ be fixed. Then for every graph $G\in \DD$ one can find a family $\Ff(G)$ of vertex subsets of $G$ with $|\Ff(G)|\leq \Oh_{\DD,p,\eps}(|G|^\eps)$ satisfying the following:
%  \begin{itemize}
%   \item for every $X\subseteq V(G)$ with $|X|\leq p$ there is $A\in \Ff(G)$ such that $X\subseteq A$; and
%   \item the class $\setof{G[A]}{G\in \DD, A\in \Ff(G)}$ has bounded shrubdepth.
%  \end{itemize}
% \end{manualtheorem}

Finally, we obtain \cref{thm:lsd-nd} (low shrubdepth covers)
%--
%stating that every structurally nowhere dense class $\DD$ of graphs admits shrubdepth covers 
%of size $\Oof_{\DD,p,\eps}(n^\eps)$ and  bounded shrubdepth, thus 
as a corollary of \Cref{thm:bush-nd}. 
This answers the question of \cite{GajarskyKNMPST20,BrianskiMPS21}.
We use an argument similar to that used in~\cite[Section~4]{GajarskyKNMPST20}.
%In this section we use \Cref{thm:bush-nd} to prove \Cref{thm:lsd-nd}. 
We first need to recall a few definitions and facts.

We need the characterization of classes of sparse graphs via low treedepth covers.
The {\em{treedepth}} of a graph $G$ is the least number $G$ satisfying the following: there exists a rooted forest $F$ of depth $d$ on the vertex set of $G$ such that for every edge $uv$ in $G$, $u$ and $v$ are bound by the ancestor relation in $F$.
A \emph{$p$-cover} of a graph $G$ is a family $\Ff(G)$ of subsets of $V(G)$
such that for every set $X \subseteq V(G)$ with $|X|\le p$ there is $A \in \Ff(G)$ with $X \subseteq A$. We will use the following standard statement saying that nowhere dense classes admit low treedepth covers of small cardinality. For the proof, see~\cite[Theorem~7.7]{sparsity}.
% 
% \begin{theorem}\label{thm:ltd-be}
% Let $\CC$ be a class of graphs with bounded expansion and $p\in \N$. Then there exists $k\in \N$ such that for every graph $G\in \CC$ one can find a $p$-cover $\Ff(G)$ of $G$ with $|\Ff(G)|\leq k$ 
% such that the class $\setof{G[A]}{G\in \CC, A\in \Ff(G)}$ has bounded treedepth.
% \end{theorem}

\begin{theorem}\label{thm:ltd-amostnd}
Let $\CC$ be an almost nowhere dense class of graphs and $p\in \N$, $\eps > 0$.
Then for every graph $G\in \CC$ one can find a $p$-cover $\Ff(G)$ with $|\Ff(G)|\leq \Oh_{\CC,p,\eps}(|G|^\eps)$
such that the class $\{G[A]\colon G\in \CC, A\in \Ff(G)\}$ has bounded treedepth.
\end{theorem}

Next, 
let us first recall the necessary facts about shrubdepth. We have defined connections models and classes of bounded shrubdepth already in~\cref{sec:intro}; see also~\cite{GanianHNOM19} for an extended discussion. We will not use those definitions directly, and instead we rely on the following fact~\cite{GanianHNOM19}.

\begin{lemma}[{\cite[Theorem 4.5]{GanianHNOM19}}]\label{lem:td-sd}
    Let $\CC$ be a class of binary structures with bounded treedepth
    and $\trans T$ be a transduction. 
    Then $\trans T(\CC)$ has bounded shrubdepth.
\end{lemma}

With $\trans T$ being the identity,
\cref{lem:td-sd} implies that every class with bounded treedepth also has bounded shrubdepth.

We will also need the following simple lemma about implementing the mechanics of quasi-bushes by a transduction. 

\begin{lemma}
\label{lem:td-quasi-bushes}
For every $d$ and finite label set $\Lambda$ there is a transduction $\trans T$ such that the following holds. Let $B$ be a quasi-bush of depth at most~$d$ that uses labels from $\Lambda$. Then $G(B)\in \trans T(B)$.
\end{lemma}
\begin{proof}
      Recall that for $u,v\in V(G)$, whether $u$ and $v$ are adjacent in $G$ can be recovered from $B$ as follows: If $w$ is the least ancestor of $v$ such that there is a pointer from $u$ to $w$, then the adjacency of $u$ and $v$ depends only on the labels of $u$ and $v$ and the label of the pointer from $u$ to $w$. It is clear that for fixed $d$ and $\ell$, this property can be expressed using a first-order formula $\varphi(x,y)$. Therefore, to obtain $G(B)$ from $B$ by means of a transduction, it suffices to apply the interpretation using $\varphi(x,y)$ and restrict the obtained graph to subgraph induced by the leaves of $B$.
     \end{proof}
     
With all the tools prepared, we can prove \Cref{thm:lsd-nd}. The the argument is similar to that used in~\cite[Section~4]{GajarskyKNMPST20}.

\begin{proof}[Proof of \Cref{thm:lsd-nd}]
Apply Theorem~\ref{thm:bush-nd} to $\DD$ to obtain $d, \ell$ and an almost nowhere dense class $\BB$ of quasi-bushes, each of depth $d$ and using a label set $\Lambda$ of size at most $\ell$. Let $G \in \DD$,
 $B \in \BB$ be a quasi-bush with $G=G(B)$, and $T$ be the tree underlying~$B$. Recall that $\leaves(T) = V(G)$. %Set $\epsilon':=\epsilon/d$.

%Since $\BB$ is slightly dense, there exists a coloring of $U(B)$ with $g(\epsilon,p)\cdot |B|^\epsilon$ colors such that each $p$-tuple of colors induces a subgraph of Gaifman graph of $B$ of treedepth at most $p$. 

Fix $p\in \N$.
Since $\BB$ is almost nowhere dense, by \Cref{thm:ltd-amostnd}, there exists a family $\EE$ of subsets of nodes of $B$ such that $|\EE|\leq \Oh_{\DD,(d+1)p,\eps}(|B|^{\eps})$ and for every set $Y$ of at most $(d+1)p$ nodes of $B$ there is $C\in \EE$ satisfying $Y\subseteq C$.

\newcommand{\anc}{\mathrm{anc}}

For a leaf of $T$, let $\anc(v)$ denote the set of all ancestors of $v$ in $T$ (including $v$); note that it is always the case that $|\anc(v)|\leq d+1$.
For a subset $Z$ of $V(T)$ and a leaf $v$ of $T$, we say that $v$ is  \emph{$Z$-closed} if $\anc(v) \subseteq Z$.
For every $C \in \EE$, let $C'$ be the set of all $C$-closed leaves of $T$. We define
\[\Ff(G) = \setof{C'}{C \in\EE}.\] 
We now verify that $\Ff=\Ff(G)$ has the desired properties.

First, clearly $|\Ff| \le |\EE|$. As $|B| \le d|G|$, we conclude that 
\[|\Ff|\leq |\EE| \leq \Oh_{\DD,(d+1)p,\eps}(|B|^{\eps})\leq \Oh_{\DD,p,\eps}(|G|^{\eps}),\] as desired.

Next, fix a $X\subseteq V(G)$ with $|X| \le p$. Let $Y = \bigcup_{v\in X} \anc(v)$. Note that $|Y|\leq (d+1)p$, so there exists $C \in \EE$ such that $Y \subseteq C$. By construction, every $v \in X$ is $C$-closed, and so every $v \in X$ also belongs to $C'$. Hence $X\subseteq C'\in \Ff$.

It remains to show that the class $\setof{G[A]}{G\in \DD, A\in \Ff}$ has bounded shrubdepth. Consider any $A\in \Ff(G)$; then $A = C'$ for some $C \in \EE$. Let $A'=\bigcup_{v\in C'} \anc(v)$; then $A'\subseteq C$ by the definition of $C'$.
From the definition of quasi-bushes it follows that whether $u$ and $v$ are adjacent in $G(B)$ can be determined from $B[\anc(u)\cup \anc(v)]$, and so the adjacency between any $u,v \in A$ can be determined from the quasi-bush $B[A']$. In other words, we have
\begin{equation}\label{eq:wydra}
G(B)[A] = G(B[A'])[A].
\end{equation}
The treedepth of $B[A']$ is not larger than that of $B[C]$, hence the class $\setof{B[A']}{G\in \DD, A\in \Ff(G)}$ has bounded treedepth. Classes of bounded treedepth also have bounded shrubdepth, and applying a fixed transduction to any class of bounded shrubdepth again yields a class of bounded shrubdepth~\cite{GanianHNOM19}. By applying this statement to the transduction $\trans T$ provided by \Cref{lem:td-quasi-bushes} for $d$ and $\Lambda$, we infer that the class $\trans T(\setof{B[A']}{G\in \DD, A\in \Ff(G)})$ has bounded shrubdepth. By~\eqref{eq:wydra} and the properties of $\trans T$, this class contains the class $\setof{G[A]}{G\in \DD, A\in \Ff}$. So the latter class also has bounded shrubdepth, as desired.
\end{proof}

% Since the Gaifman graph $G_{B[C]}$ of $B[C]$ has treedepth at most $d$, by Lemma~\ref{lem:td-quasi-bushes} 
% one can obtain $G[A]$ from $G_{B[C]}$ by a fixed transduction $\trans T$, as this implies that $G[A]$ comes from a fixed graph class of bounded shrubdepth depending on $d$ and $T$.

\bibliography{references}{}
\bibliographystyle{plain}

\appendix
%!TEX root = main.tex

\section{Proofs from \cref{sec:bushes}}

    The remainder of this section is devoted to the proof of \cref{thm:trandsuction-thm}, stated in \cref{sec:transduction section}.
    That is, we show that there is a transduction $\trans B$ 
that given a graph $G\in\CC$ outputs the bush $B_G\coloneqq B(G,\le)$ as constructed 
in \cref{sec:bush-construction},
for a suitable order $\le$ on $G$. Note that the order is not given on input to the transduction.
As a side effect, we obtain a different 
proof that the class $\setof{B_G}{G\in \CC}$ has bounded expansion,
which does not rely on the existence of universal orders
(\cref{cor:universalBE}) 
but instead relies on the results of 
\cite{GajarskyKNMPST20} (precisely, \cref{thm:lsd-be}).

Instead of proving \cref{thm:trandsuction-thm} directly, we prove the following, weaker statement.
Recall that a graph $G$ is \emph{$k$-degenerate} if each of its subgraphs contains a vertex of degree at most $k$. 

%It is well-known that $G$ is $k$-degenerate if and only if its edges can be oriented so that the resulting directed graph is acyclic (has no directed cycles),
%and every vertex has out-degree at most~$k$.

\begin{lemma}\label{lem:transduction-lemma}
  Fix a class $\CC$ of graphs with bounded expansion 
  and a formula $\phi(x,y)$.
  Then there is a class of bushes $\BB$ of bounded depth 
  and using a bounded number of labels, and 
  a transduction $\trans B\from \CC\tto \BB$,
  such that for every $G\in \CC$ there is a bush $B\in\trans B(G)$ representing $\trans I_\phi(G)$.
  Moreover, the Gaifman graphs of bushes in $\BB$ have bounded degeneracy.
\end{lemma}

 Using results of \cite{GajarskyKNMPST20}, we first show how \cref{thm:trandsuction-thm} can be obtained from \cref{lem:transduction-lemma}.
\begin{proof}[Proof of \cref{thm:trandsuction-thm} using \cref{lem:transduction-lemma}] 
  Let $\DD$ be a class with structurally bounded expansion,
  and let $\CC$ and $\trans T\from \CC\tto \DD$ and $\trans T'\from \DD\tto \CC$ be as in \cref{thm:sparsification},
  so that $H\in \trans T(\trans T'(H))$ for all $H\in\DD$
  and $\trans T$ is non-copying. By adding additional colors to graphs in $\CC$ we may assumed that $\trans T$ is an interpretation. Let $\phi(x,y)$ be the formula underlying it;
  then $\DD$ is contained in the hereditary closure of $\trans I_\phi(\CC)$.

  Let $\trans B\from \CC\tto \BB$ be the transduction 
  from \cref{lem:transduction-lemma}.
  By composition, we get a transduction $\trans B\circ \trans T'\from \DD\tto \BB$ such that 
  for every $H\in \DD$ there is $G\in\trans T'(H)$ such that
  $H\in \trans T(G)$, and 
  there is a bush  $B\in (\trans B\circ \trans T')(H)$ representing $\trans I_\phi(G)$.
  Note that $H\in\trans T(G)$ implies 
  that $H$ is an induced subgraph of $\trans I_\phi(G)$, by the definition of $\phi$.

  Let $\trans B'\from \DD\tto \BB'$ be the transduction $\trans B\circ \trans T'$ followed by a transduction that, given a bush $B$, first selects an arbitrary subset of leaves $W$ and then restricts $B$ to $B[W]$. (We defined the restriction of bushes to leaf subsets in the proof of \cref{lem:restrict}.) It follows from the above that $\trans B'\from \DD\tto \BB'$ 
  is a transduction such that $\trans B'(H)$ contains 
  some bush representing $H$, for every $H\in \DD$. Let $\BB' \coloneqq \trans B'(\DD)$. Then $\BB'$ is a class of bushes of bounded depth and using a bounded number of labels,
  and the class of Gaifman graphs of bushes in $\BB'$ has bounded degeneracy. Moreover, $\BB'$ has structurally bounded expansion, as it can be transduced from the bounded expansion class $\CC$ using $\trans B'\circ \trans T$.
  
As argued in \cite[Section 5.3]{GajarskyKNMPST20}, \cref{thm:lsd-be} implies the following: every class that has structurally bounded expansion and bounded degeneracy actually has bounded expansion. Hence, the class of Gaifman graphs of structures in $\BB'$ has bounded expansion.
This proves \cref{thm:trandsuction-thm}. 
\end{proof}

Before proving \cref{lem:transduction-lemma}, 
 we prove an auxiliary lemma about transducibility of the weak reachability relation. The main idea is to inductively use \cref{lem:direction transduction}.
 
 %to simulate the mechanism of transitive fraternal augmentations (see~\cite[Section~7.4]{sparsity}).

\begin{lemma}\label{lem:wreach transduction}
  Fix a number $r\in\N$.
  Let $\CC$ be a graph class with bounded expansion, and for each $G\in\CC$ fix an order $\le$ 
  such that $\wcol_{2r}(G,\le)$ is bounded by some constant $d$.
Then there is a transduction $\trans W_r$ that given a graph $G\in\CC$
outputs the set $V(G)$ endowed with the binary relation 
$W_r$ consisting of pairs $(u,v)$ such that ${v\in \wreach_r[u]}$.
\end{lemma}
\begin{proof}
    For every $s\in \{0,1,\dots,r\}$, let $W_s \coloneqq \setof{(u,v)}{v\in \wreach_s[u]}$ and let $G_s$ be the undirected graph underlying the directed graph $(V(G),W_s)$. 
    
    We first observe that $G_r$ has star chromatic number at most $d+1$. To see this, color $G$ with $d+1$ colors so that whenever $u\in \wreach_{2r}[v]$, $u$ and $v$ receive different colors; this can be done by a greedy left-to-right coloring using the assumption that $\wcol_{2r}(G,\le)\leq d$. It is then easy to verify that this coloring, call it $\rho$, is a star coloring of $G_r$. (This is a standard construction.) Indeed, observe that if $u,v,v'$ are three distinct vertices such that $v<u$ and $uv,uv'$ are edges in $G_r$, then all these three vertices need to receive three different colors in $\rho$, because in every pair of them one is weakly $2r$-reachable from the other (see  \cref{lem:ordering wreach}). It follows that if $C$ is any connected subgraph of $G_r$ receiving at most two colors in total, then $C$ must be a star with the center being the $\leq$-smallest vertex. So indeed, $\rho$ is a star coloring of $G_r$. Since $G_s$ is a subgraph of $G_r$ whenever $s\leq r$, it follows that each $G_s$ also has star chromatic number at most $d+1$.

    By induction on $s\in \{0,1,\ldots,r\}$, we construct transductions $\trans W_s$ that given $G$, output the set $V(G)$ endowed with the relation $W_s=\setof{(u,v)}{v\in \wreach_s[u]}$. 
    In the base case, when $s=0$, the transduction $\trans W_0$ outputs the identity relation.
    
    In the inductive step, suppose $s\ge 1$ and the transduction $\trans W_{s-1}$ has already been constructed. We then construct~$\trans W_{s}$.
  
    Let $G\in\CC$.
    For each $u \in V(G)$ and every $1\le i<|\wreach_s[u]|$
    fix a path $\pi_{u}^i$ of length at most $r$ that starts at $u$ and ends at the $i$th $\le$-smallest vertex $w$ of $\wreach_s[u]$
     such that $w$ is the $\le$-smallest vertex of $\pi_{u}^i$.
    Then let $f_i(u)$ be the second vertex on the path $\pi_u^i$. Thus, $\setof{f_i}{1\leq i\leq d}$ are partial functions from $V(G)$ to $V(G)$ such that, for every $u\in V(G)$, if $f_i(u)$ is defined, then $uf_i(u)$ is an edge in $G$. Hence, we may apply Lemma~\ref{lem:direction transduction} to obtain, for each $1\leq i\leq d$, a transduction $\trans Q_i$ that, given $G$, computes (as one of the possible outputs) the partial function $f_i$. 
    
   For each $1\le j\le d$, let $g_j(u)$ denote the $j$th $\le$-smallest vertex in $\wreach_{s-1}[u]$, provided it exists. As the partial function $g_j$ is contained in the relation $W_{s-1}$, by the inductive assumption and again \cref{lem:direction transduction} (but this time applied to $G_{s-1}$),
   there is a transduction $\trans R_j$ that, given $G$, computes (as one of the possible outputs) the partial function $g_j$.
  
  Now the key observation is the following: if it holds that $v\in \wreach_s[u]$, then 
  $v\in\wreach_{s-1}[u]$ or 
  there are some $1\le i,j \le d$ for which there is a (unique) $w'\in V(G)$ such that 
  \begin{itemize}[nosep]
    \item $w'=f_i(u)$, and
    \item $v=g_j(w')$.
  \end{itemize}
  For every vertex $v\in \wreach_{s}[u] \setminus \wreach_{s-1}[u]$, 
  pick any pair $(i,j)$ for which there is $w'$ satisfying the above conditions, and collect those pairs in a set $C(u)$.
  
  We may now define the transduction $\trans W_s$. Given a graph $G$, the transduction $\trans W_s$ first colors each vertex $u$ by $C(u)$,
  and then produces pairs $(u,v)$ for which $v \in \wreach_{s-1}[u]$, or there is some pair $(i,j)\in C(u)$ and 
  some vertex $w'$ such that $u,v,w'$ satisfy the conditions stated above. These requirements can be checked using the relation $W_{s-1}$, the functions $f_i$, and the functions $g_j$, introduced using the transductions $\trans W_{s-1}$, 
  $\trans Q_i$, and~$\trans R_j$, respectively. This finishes the induction step.
  
  Once all transductions $\trans W_0,\trans W_1,\dots,\trans W_r$ are constructed, we note that $\trans W_r$ satisfies the conclusion of the lemma.
  \end{proof}

In the rest of this section, we prove \cref{lem:transduction-lemma}. 
So let $\cal C$ be a class of (colored) graphs of bounded expansion and let $\phi(x,y)$ be a formula.
Let $r,q$ be obtained by applying \cref{lem:separation and types} to $\phi$.
Let $d$ be such that $\wcol_{4r}(G)$ is at most $d$, for all $G\in \CC$. 
%This strengthens the assumption in \cref{sec:common-notation}, by bounding $\wcol_{2r}$ instead of just $\wcol_{r}$.

Fix $G\in\CC$ and an order $\le$ on $V(G)$ such that  ${\wcol_{4r}(G,\le)}$ is at most $d$.
Consider the bush $B(G,\le)$ constructed in \cref{sec:bush-construction} for this particular choice of the order $\le$ and radius parameter $r$. (Note that we use $r$ in the construction of the bush, but assume boundedness of weak reachability set for distance $4r$.)
We define an auxiliary directed graph $D$ as follows: 
\begin{itemize}[nosep]
 \item The vertex set of $D$ is $V(B(G, \le))$. Recall that this means that the vertices of $D$ are tuples $\first_i(v)$ for all $0\le i\le d$ and $v\in V(G)$.
 \item For every pair of distinct nodes $X,Y$ of $D$, insert a directed edge $(Y,X)$ to $D$ if $X=\emptyset$ or $\max X\in Y$. Note that we do not require $X$ and $Y$ to be at the same depth in $B$.
\end{itemize}
Note that the Gaifman graph of the bush $B(G,\le)$
is a subgraph of the undirected graph underlying $D$. 
%Moreover, the directed graph $D$ is acyclic.

We observe that $D$ has bounded maximum outdegree.

\begin{lemma}\label{lem:degeneracy-D}
  Every vertex of $D$ has out-degree at most ${1+d^2\cdot 2^{d}}$.
\end{lemma}
\begin{proof}
Fix a vertex $Y$ of $D$; we would like to bound the number of vertices $X$ for which the edge $(Y,X)$ is present in $D$. By putting $X=\emptyset$ aside and accounting for a $+1$ summand in the final bound, we may assume that $X$ is nonempty. Then, by \cref{lem:ordering wreach},
we have $X\subset \wreach_{2r}[\max X]$.
The existence of the edge $(Y,X)$ implies that $\max(X)\in Y$, so $X\subseteq \wreach_{2r}[w]$ for some $w\in Y$. For any fixed $w\in Y$ there are
only at most $2^{d}$ many subsets of $\wreach_{2r}[w]$, and due to padding in the definition of tuples $\first_i(\cdot)$, each of them can give rise to at most $d$ vertices of $D$. Hence, there are at most $d^2\cdot 2^d$ nonempty tuples $X$ for which the edge $(Y,X)$ is present.
\end{proof}

It is well-known that if a graph $H$ has an orientation in which all out-degrees are bounded by $k$, then $H$ is $2k$-degenerate. Therefore, from \ref{lem:degeneracy-D} it follows that the undirected graph underlying $D$ is $(2+d^2\cdot 2^{d+1})$-degenerate. So the Gaifman graph of $B(G,\le)$ is also $(2+d^2\cdot 2^{d+1})$-degenerate.

Next, we verify that $D$ can be produced by a transduction.

\begin{lemma}\label{lem:trans-for-directed}
  There is a transduction $\trans B'$ that, given $G\in \CC$, outputs (as one of possible outputs) the directed graph $D$ defined above.
\end{lemma}
\begin{proof}
Given $G$, the transduction $\trans B'$ first produces $d+1$ copies of $V(G)$,
producing the set $V(G)\times \set{0,\ldots,d}$.

Note that the relation $u\in \wreach_r[v]$ can be obtained by applying the transduction $\trans W_r$
constructed in \cref{lem:wreach transduction}.
Moreover, 
by \cref{lem:ordering wreach}, the order $\le$ and the relation $W_{2r}=\setof{(u,u')}{u' \in \wreach_{2r}(u)}$
agree on the set $\setof{u}{u\in \wreach_r[v]}$, for any given $v$.
Therefore, using the transductions $\trans W_{r}$ and $\trans W_{2r}$
constructed in \cref{lem:wreach transduction}, we can obtain the relation $\setof{(u,v)}{u\in \wreach_r[v]}$ and the intersection of the order $\le$ with the set $\bigcup_{v\in V(G)} \wreach_r[v]\times \wreach_r[v]$.

Next,
the transduction $\trans B'$ creates an edge from $(v,i)$ to $(w,j)$
if $\max(\first_i(v))\in \first_j(w)$ or $j=0$, which can be defined using the objects provided above.
Finally, $\trans B$ restricts the domain to any maximal subset $U$ of 
$V(G)\times \set{0,\ldots,d}$ such that $\first_i(v)\neq\first_i(w)$, for all distinct $(v,i),(w,i)\in U$. This defines the directed graph $D$ that is output by~$\trans B'$.
\end{proof}

To produce the bush $B(G,\le)$,
recall that the Gaifman graph of $B(G,\le)$ is a subgraph of the undirected graph underlying $D$. Hence, $B(G,\le)$ can be obtained from $D$ using the transduction $\trans B''$ given by~\cref{lem:direction transduction}.
%That is, there is a transduction $\trans B$ such that $B(G,\le)\in \trans B(G)$, for every $G\in \CC$.
Composing this transduction with the transduction $\trans B'$ of \cref{lem:trans-for-directed} yields a
transduction $\trans B$ such that $B(G,\le) \in \trans B(G)$ holds for every $G \in \cal C$.

Let $\BB\subseteq \trans B(\CC)$ be the family comprising all bushes contained in $\trans B(\CC)$ whose Gaifman graph has degeneracy at most $2+d^2\cdot 2^{d+1}$. Then $\BB$ is a class with structurally bounded expansion by definition, and with the underlying Gaifman graphs having bounded degeneracy, also by definition. And as argued, for every $G\in \CC$ there is an order $\le$ on $V(G)$ such that $B(G,\le)\in \BB$. 
So this finishes the proof of \cref{lem:transduction-lemma},
and hence of \cref{thm:trandsuction-thm}.

\end{document}